\documentclass{aa}
\usepackage{natbib,epsfig,txfonts,graphicx}

\def\Ha{H$\alpha$}

\def\Hb{H$\beta$}

\def\Hg{H$\gamma$}

\def\HI{H\,{\sc i}}

\def\HeI{He\,{\sc i}}
\def\HeII{He\,{\sc ii}}
\def\HeIII{He\,{\sc iii}}
\def\SiII{Si\,{\sc ii}}
\def\SiIII{Si\,{\sc iii}}
\def\SiIV{Si\,{\sc iv}}
\def\SIV{S\,{\sc iv}}

\def\lg{\log g}

\def\logQ{\log Q}

\def\Teff{T_{\rm\kern-.15em ef\kern-.05em f}} 
\def\Mdot{\dot M}
\def\Msun{M_\odot}
\def\Rsun{R_\odot}

\def\yhe{Y_{\rm He}}

\def\Rstar{R_{\ast}}
\def\Mstar{M_{\ast}}
\def\Lstar{L_{\ast}}

\def\vinf{v_{\rm \infty}}
\def\vesc{v_{\rm esc}}
\def\vsini{v\,\sin i}

\def\vmicro{v_{\rm mic}}
\def\fcl{f_{\rm cl}}
\def\Dmom{D_{\rm mom}}

\def\Fb{F^{\rm neb}_\beta}
\def\Fn{F^{\rm neb}_{\lambda}}
\def\Fs{F^\ast_{\lambda}}

\def\lam{$\lambda$}

\def\kms{\mbox{km\,s}^{-1}}

\def\beq{\begin{equation}}
\def\eeq{\end{equation}}
\def\beqa{\begin{eqnarray}}
\def\eeqa{\end{eqnarray}}

\def\Htoe{\mbox{He 2$-$108}}
\def\Htts{\mbox{He 2$-$260}}
\def\Mots{\mbox{M 1$-$37}}
\def\Mote{\mbox{M 1$-$38}}
\def\Mtot{\mbox{M 2$-$12}}
\def\Mttt{\mbox{M 2$-$33}}

\begin{document}

\title{Central stars of planetary nebulae in the Galactic bulge\thanks{The
data presented herein were obtained at the W.~M.~Keck Observatory, which is
operated as a scientific partnership among the California Institute of
Technology, the University of California and the National Aeronautics and
Space Administration. The Observatory was made possible by the generous
financial support of the W.~M.~Keck Foundation.}}
\author{P.~J.~N.~Hultzsch\inst{1}, J.~Puls\inst{1}, R.~H.~M\'{e}ndez\inst{2},
A.~W.~A.~Pauldrach\inst{1}, R.-P.~Kudritzki\inst{2}, T.~L.~Hoffmann\inst{1},
J.~K.~McCarthy}

\offprints{P.~Hultzsch}

\institute{Universit\"{a}ts-Sternwarte M\"{u}nchen (USM), Scheinerstr.~1, 81679
M\"{u}nchen, Germany \and Institute for Astronomy, University of Hawaii, 2680
Woodlawn Drive, Honolulu, HI 96822, USA}

\date{20/10/2006 ; 27/02/2007 }

\abstract {Optical high-resolution spectra of five central stars of
planetary nebulae (CSPN) in the Galactic bulge have been obtained with
Keck/HIRES in order to derive their parameters. Since the distance of the
objects is quite well known, such a method has the advantage that stellar
luminosities and masses can in principle be determined without relying on
theoretical relations between both quantities.}
{By alternatively combining the results of our spectroscopic investigation
with evolutionary tracks, we obtain so-called spectroscopic distances,
which can be compared with the known (average) distance of the bulge-CSPN.
This offers the possibility to test the validity of model atmospheres and
present date post-AGB evolution.}
{We analyze optical H/He profiles of five Galactic bulge CSPN (plus one
comparison object) by means of profile fitting based on state of the art
non-LTE modeling tools, to constrain their basic atmospheric parameters
($\Teff$, $\lg$, helium abundance and wind strength). Masses and other
stellar radius dependent quantities are obtained from both the known
distances and from evolutionary tracks, and the results from both approaches
are compared.}
{The major result of the present investigation is that the derived
spectroscopic distances depend crucially on the applied reddening law.
Assuming either standard reddening or values based on radio-\Hb\ extinctions,
we find a mean distance of $9.0 \pm 1.6$\,kpc and $12.2 \pm 2.1$\,kpc,
respectively. An ``average extinction law'' leads to a distance of $10.7
\pm 1.2$\,kpc, which is still considerably larger than the Galactic center
distance of 8\,kpc. In all cases, however, we find a remarkable {{\it
internal agreement}} of the individual spectroscopic distances of our sample
objects, within $\pm 10$\% to $\pm 15$\% for the different reddening laws.}
{Due to the uncertain reddening correction, the analysis presented here
cannot yet be regarded as a consistency check for our method, and a rigorous
test of the CSPN evolution theory becomes only possible if this problem
has been solved.}


\titlerunning{CSPN in the Galactic bulge}
\authorrunning{P.~Hultzsch et al.}

\maketitle

\section{Introduction}
\label{intro}

Planetary nebulae (PN) in the direction of our Galactic bulge attract
much interest, in particular because of their abundances.  They are also
important, however, because the Galactic bulge is one of the places in
the universe with a reasonably well-known distance; for that reason it
is a natural ground for testing methods of PN distance determination. Our
purpose in this paper is to test so-called spectroscopic distances, i.e.,
distances obtained from a determination of the basic atmospheric parameters
of PN central stars ($\Teff$, $\lg$) using model atmospheres to interpret
the observed central star spectra.

An alternative idea is to take good spectra of PN central stars in
the Magellanic Clouds which until now has only been done in the far-UV
\citep{Herald04}. To perform such a campaign in the optical has to await
the advent of new technologies and instrumentation, e.g., adaptive optics
at optical wavelengths and/or space-born faint object high resolution
spectroscopy. High spectral resolution is necessary, because unlike massive O
stars, PN central star spectra are contaminated by strong nebular emissions
affecting the main H and He diagnostic absorption lines. In view of this,
the Galactic bulge has the essential advantage over the Magellanic Clouds
that it is much closer to us. It also has a disadvantage: the interstellar
extinction is much higher towards our bulge, and we do not fully understand
the properties of the interstellar medium in that direction. We refer the
reader to the study by \cite{Stasinska92}, who compared Balmer decrement
extinctions versus radio-H$\beta$ extinctions and concluded that for most
distant PN the ratio $R$ of total to selective extinction must be lower
than the canonical value of 3.1, and suggested a value of \mbox{$R$ = 2.7}.

However, for the moment the advantage of a smaller distance more than
compensates for the extinction disadvantage, and so we started a project
made possible by the advent of the high-resolution spectrograph HIRES at
the Keck 10-m telescope (Mauna Kea, Hawaii).  Spectra of several bulge
central stars were taken from 1994 to 1996, and a preliminary description
of the best of these spectra was presented by \cite{Kudritzki97}.

The 10-year delay between observations and the present paper can be
attributed to the lack of confidence in the last generation of model
atmospheres.  However, plane-parallel, metal-free non-LTE models are no
longer the state of the art, and nowadays we are able to produce much
more sophisticated models that take into account blocking and blanketing
by numerous metal lines, and feature hydrodynamically plausible winds in
an extended, expanding atmosphere. The analysis of the objects performed
in the present paper is based on this new generation of model atmospheres,
and it turned out that the determination of stellar atmospheric parameters
(e.g.  $\Teff$ and $\lg$) appears to be robust enough for spectroscopic
distance determinations.

Spectroscopic distances as determined here are not independent of stellar
evolution theory; we can make the jump from $\Teff$ and $\lg$ to mass
and distance because we assume that the relation between core mass and
luminosity, and the post-AGB evolutionary tracks, are correct.  In summary,
what we are testing here is whether or not the combination of post-AGB
evolution theory and the presently best available model atmospheres can
successfully predict the distance to the Galactic bulge from high-resolution
optical spectra of a sample of five bulge PN central stars.  We present
a rigorous error analysis for this test.

\smallskip \noindent This paper is organized as follows: In
Section~\ref{selection} we comment on the selection criteria for our
sample, the observations and data reduction. Section~\ref{det} comprises
a brief summary of the NLTE atmosphere code used for the analysis, the
method applied to derive stellar and wind parameters from the spectra
and some comments on the profile fits and spectroscopic results for the
individual objects of our sample.
In Section~\ref{deduced} we investigate those parameters which are not
{\it directly} measurable from the spectra (particularly masses and radii),
by introducing two different analysis methods, based either on evolutionary
tracks or adopted distances. We compare the results, with particular emphasis
on the involved error sources and the obtained degree of precision. We
present the spectroscopic distances, and investigate the wind-momentum
luminosity relation for our sample. In Section~\ref{discussion}, we
discuss the results and comment on several recent developments like the
new trigonometric parallaxes by \citet{Harris07}. Section~\ref{summary}
summarizes our conclusions and provides future perspectives.

\section{Object selection and observations}
\label{selection}

The PN central stars (CSPN) to be observed were selected among the brightest
in the direction towards the Galactic center, according to the following
criteria:
\begin{itemize}
\item[(1)] located within 10 degrees of the direction to the Galactic center;
\item[(2)] angular size smaller than 10 arc sec, to reduce the
probability of foreground PN;
\item[(3)] only low-excitation PN, to ensure low $\Teff$, because in that
way we know the stars are evolving at high luminosities towards higher
$\Teff$; in addition, at low enough $\Teff$ it is easier to determine
$\Teff$ from the ionization equilibrium of \HeI\ and \HeII\ stellar features.
\end{itemize}
For comparison with the previously studied ``solar neighborhood''
CSPN, we selected the central star of \Htoe.  We have chosen this
particular object because it has been analyzed by different groups and,
more importantly, using different methods. Previous analyses performed
in a similar way as our approach resulted in a spectroscopic distance
of 6.0\,kpc (\citealt{Mendez92} , see also \citealt{Kudritzki97} and
2006), whereas alternative approaches were either consistent with these
values \citep{Napiwotzki06} or gave considerably different results
\citep{Pauldrach04} (see Section~\ref{discussion}).

The bulge CSPN spectrograms were obtained with the 10\,m Keck telescope,
Mauna Kea, Hawaii, using the high-resolution echelle spectrograph HIRES
\citep{Vogt94}. The spectra cover the range from 4250\,\AA\ to 6750\,\AA\
with a spectral resolution of 0.1\,\AA. Individual exposures were between
100\,s and 1800\,s, and several exposures were made for each of the objects,
cf.\ Table~\ref{obslog}. In order to avoid excessive contamination of
one order by neighboring ones, particularly at shorter wavelengths, we
selected the rather short slit ``C5'' ($7.0 \times 1.148$\,arcsec) for
our observations. This ensured a clearly defined inter-order minimum and
reduced the effect of scattered light from strong nebular lines, which
produced a halo affecting in some cases more than one echelle order.

The CCD image contained 31 echelle orders, spread over $2048\times 1024$
pixels. During each observing night dark frames, flat fields, and Th-Ar
comparison spectra were obtained immediately before and\,/\,or after the
object exposure.  The spectrograms of the central star of \Htoe\ were
obtained with the 3.6\,m telescope and Cassegrain echelle spectrograph
(CASPEC) of the European Southern Observatory, La Silla, Chile. For a
description of the reductions we refer to \citet{Mendez88}. From these
spectrograms we used \Hb, \HeI\ \lam \lam 4471, 4387, 4922 and \HeII\ \lam
4200.

Additional spectra of the central star of \Htoe\ were taken with the 3.5\,m
ESO New Technology Telescope (NTT) and EMMI Cassegrain spectrograph at the
Cerro La Silla, Chile, in March 1994. From these spectrograms we used \Hg,
\Ha, \HeI\ \lam 4713 and \HeII\ \lam \lam 4686 and 4541.

The usual calibration exposures (flat field, comparison spectra) were
obtained. For brevity we will not explain the reduction of these simple
spectra but will concentrate on the echelle spectra only.

HIRES spectra reduction was performed with IRAF,\footnote{IRAF is
distributed by the National Optical Astronomy Observatories, which are
operated by the Association of Universities for Research in Astronomy,
Inc., under cooperative agreement with the National Science Foundation.}
using the following procedure: First we removed hot pixels and cosmic rays
and subtracted the dark frames. Thereafter, the run of the order maxima
was fitted, and a pixel shift between object and flat field frames was
identified. Next, the object images as well as the comparison spectra were
corrected for the normalized flat field images and the individual echelle
orders of the object spectra were extracted according to the fitting
functions determined previously. Then the wavelength scale was calibrated
using the comparison lines, which ensured an uncertainty of about 0.02\,\AA.

Finally, the extracted spectra were rectified. This process turned out
to be quite delicate because of the high resolution of the spectra and
the contamination by nebular light. Due to the high resolution, some of
the broader stellar lines like \Ha\ were spread out over more than one
echelle order, which made the identification of the stellar continuum in
this range rather difficult. This challenge was met by interpolating the
continuum run of neighboring echelle orders. With this method, we were
able to rectify all echelle orders.

\begin{table}
\begin{center}
\caption{Summary of observations (Galactic bulge objects).}
\renewcommand{\arraystretch} {1.1}
\begin{tabular*}{\columnwidth}{l@{\extracolsep\fill}c|crc}
\hline \hline
&& number of & \multicolumn{2}{c}{total exposure}\rule{0cm}{3ex}\\
&~~object & observations & \multicolumn{2}{c}{time (seconds)} \\[.2ex]
\hline
&\Htts      & 9 &  ~~~~~~~~~~~~~~~~6700&\rule{0cm}{3ex}\\
&\Mots      & 6 & 12000& \\
&\Mote      & 2 &  3300& \\
&\Mtot      & 5 &  5500& \\
&\Mttt      & 9 &  8080& \\[.2ex]
\hline \hline
\end{tabular*}
\label{obslog}
\end{center}
\end{table}

The light of the nebula not only contaminates the stellar line profiles but
also the stellar continuum. Therefore, an adequate correction is needed in
order to obtain the real stellar line profiles, which (in case of absorption)
become stronger in relation to the normalized continuum after accounting
for the nebular continuum. When exposing the spectrograms, the nebulae
were not fully covering the slit, and the insufficient spatial resolution
of the nebulae prevented a simple subtraction. To correct for the nebular
continuum passing through the slit, we applied the following procedure:

\citet[ p. 84]{Pottasch83} showed that the ratio $R_\beta$ of the integrated
flux $\Fb$ in the \Hb\ line to the nebular continuum, $\Fn$,
\begin{equation}
 R_\beta = \frac{\Fb}{\Fn} \approx \rm const
\end{equation}
only weakly varies as a function of the temperature and electron density in
the planetary nebula, as long as the considered range is not too large. We
adopt a representative value of $R_\beta = 2000$\,\AA, as we expect the
nebular electron temperatures to be relatively low, due to the low
temperatures (28\,000\,K to 39\,000\,K) of our central stars. Since the
spectrograms have already been normalized for stellar plus nebular continuum,
the correction for nebular continuum must be performed using the equivalent
width $W_\beta$ of the nebular \Hb\ line instead of the flux~$\Fb$. The
corrected flux of the stellar continuum then follows as
\begin{equation}
 \Fs = 1 - \frac{W_\beta}{R_\beta}\,,
\end{equation}
which, again, had to be normalized after the correction.

\begin{table}[b!]
\begin{center}
\caption{Average values of $W_\beta/R_\beta$, required to correct for nebular
continuum contamination (see text).}
\tabcolsep=4pt
\renewcommand{\arraystretch} {1.3}
\begin{tabular*}{\columnwidth}{l@{\extracolsep\fill}|ccccc}
\hline \hline
object &\Htts & \Mots & \Mote & \Mtot & \Mttt\\
\hline
$W_\beta/R_\beta ~~~ $ & 0.037 & 0.020 & 0.032 & 0.030 & 0.042 \\
\hline \hline
\end{tabular*}
\label{nebcont}
\end{center}
\end{table}

A similar calculation can be made using the \Ha\ and \Hg\ emission
lines, if we adopt the Balmer Decrement.  For cool nebulae we have
$\mbox{\Hg}/\mbox{\Hb}=0.47$ and $\mbox{\Ha}/\mbox{\Hb}=2.86$\,
\citep{Osterbrock89}. From this and the measured equivalent widths of the
three emission lines, we found the average values for the correction term,
$W_\beta/R_\beta$ (see Table~\ref{nebcont}).

\subsection{Spectral morphology}

The central stars in our sample turned out to be all of the H-rich variety,
i.e., with clearly visible H Balmer lines, implying a roughly normal He
abundance, and describable in a first approximation as late-O-type spectra,
with the only exception of \Mttt, which has a somewhat hotter central
star, O5f(H) (for a general attempt at central star classification, see
\citealt{Mendez91} and references therein).

In addition, it was immediately apparent that these stars show different
degrees of wind contamination. \Htts, \Mtot\ and \Mote\ show O central
star spectra dominated by stellar absorptions, while \Mttt\ shows \HeII\
\lam 4686 in emission, traditionally classified as Of.

\begin{figure}
\resizebox{\hsize}{!}{\includegraphics{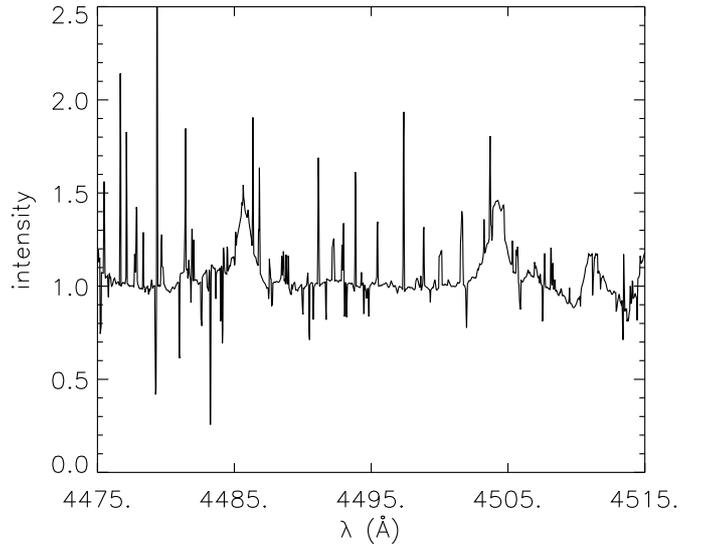}}
\caption{Famous emission lines at \lam\lam 4485, 4504 in the spectrum
of \Mots\, (see text).}
\label{m137un}
\end{figure}

Finally, we have the spectacular spectrum of \Mots, with rather weak \HeII\
\lam 4541 in absorption and \HeII\ \lam 4686 in emission, but with impressive
P Cygni profiles in \HeI\ \lam\lam 4471, 4922, 4387, 4713, and especially
in \SiIII\ \lam\lam 4552, 4568, and 4575 (see Figure~\ref{m137Si}). The
spectrum of this central star is unique and we are forced to classify it
as peculiar.  In fact, the \SiIII\ lines are so strong that one might
be tempted to classify this star as a low-gravity B0 or B2 (see, e.g.,
\citealt{Walborn90}), which is of course impossible since \HeII\ features
are present. Strong \SiIII\ at higher temperatures is already a hint of
a strong Si overabundance or at least of very specific, \SiIII\ enhancing
atmospheric properties.

Another interesting detail is the presence of very strong emission lines at
\lam\lam 4485, 4504 (Figure~\ref{m137un}). These famous lines, frequently
seen in late-O supergiants and CSPNs (see, e.g., \citealt{Swings40},
\citealt{Wolff63}, \citealt{Heap77} and \citealt{Thackeray77}), have been
attributed to \SIV\ by \cite{Werner01}.

\section{Determination of stellar and wind parameters}
\label{det}

\subsection{The model atmosphere}

\label{modelatm}

For the determination of the atmospheric and wind parameters of these
stars we use the non-LTE, spherically symmetric model atmosphere code {\sc
fastwind} (Fast Analysis of STellar atmospheres with WINDs), which was first
introduced by \citet{Santolaya97}.  Since then many improvements have been
implemented, most notably the inclusion of line blocking/blanketing effects
and the calculation of a consistent temperature structure by exploiting
the condition of flux-conservation in the inner part of the atmosphere and
the electron thermal balance (e.g., \citealt{Kubat99}) in the outer part of
the atmosphere. An in-depth explanation of the methods used in the code as
well as a comparison of representative results with those from alternative
NLTE codes {\sc cmfgen} \citep{Hillier98} and WM-basic \citep{Pauldrach01}
can be found in \citet{Puls05}.

A number of spectroscopic investigations of CSPN have been performed using
{\sc fastwind} without blanketing (e.g., \citealt{Kudritzki97}) and with
blanketing and wind clumping \citep{Kudritzki06}, where the possibility of
including wind clumping in the atmospheric treatment has been incorporated
(and carefully tested) recently. Our approach follows closely the original
approach by \citet{Schmutz95}, i.e., we assume that clumping is a matter of
{\it small-scale} density inhomogeneities in the wind\footnote{in contrast to
large-scale inhomogeneities such as co-rotating interaction zones (e.g.,
\citealt{Cranmer96} and references therein).}, which redistribute the
matter into clumps of enhanced density embedded in a rarefied, almost
void medium.  The amount of clumping is conveniently quantified by
the so-called clumping factor, $\fcl \ge 1$, which is a measure of the
over-density inside the clumps (compared to a smooth flow of identical
average mass-loss rate).\footnote{An alternative description is based on
the volume-filling factor, $f_{\rm V} = \fcl^{-1}$.} Remember that all
processes that are controlled by opacities depending on the {\it square}
of the density (such as hydrogen recombination lines and bf/ff continua
in hot stars) are preserved as long as the {\it apparent} mass-loss rate,
$\Mdot \sqrt{\fcl}$, remains constant, irrespective of the individual
values of $\Mdot$ and $\fcl$ themselves (at least as long as $\fcl$ is
spatially constant throughout the corresponding line/continuum forming flow).

In the present analysis we considered the possibility of clumping for
some of the objects (namely \Htoe, \Mote, \Mtot\ and \Mttt), in order to
explain certain inconsistencies between \Ha\ and \HeII\ 4686 arising in
unclumped models (see Section~\ref{clumping}).

In a first approach, we used a rather simple stratification (but see
\citealt{Puls06}) for the clumping factor, allowing for an unclumped lower
atmosphere (typically until $r = 1.01\,\Rstar$, where $\Rstar$ is located at
$\tau_{\rm Ross} = 2/3$), and a linear increase of the clumping factor until
we reach its maximum (fit-) value ($\fcl = 7$...$30$) at $r = 1.1\,\Rstar$
(corresponding to typically 10\% of $\vinf$), assuming that the wind-line
forming region is uniformly clumped. In this way, we account for the fact
that any instability (in particular the line-driven one, which is generally
considered as the driver for wind clumping) needs a certain time to evolve
and to become non-linear (e.g., \citealt{Runacres02}). Although there are
observational indications (from UV-spectroscopy) that clumping might start
from the wind-base on \citep{Hillier03, Bouret05}, so far we have found
no problems with our simplified approach.

\subsection{Spectral analysis}
\label{specan}

Projected equatorial rotational velocities \,$\vsini$\, (in the range
from \mbox{50\,$\kms$ to 100\,$\kms$)} were determined from the He-lines
while fitting the theoretical profiles to the observed ones. Although
rotation might not be the only effect for line broadening\footnote{E.g.,
macro-turbulence was found to be significant at least in B-type supergiants
\citep{Ryans02}, but cannot be determined here due to missing weak
metallic lines}, the values derived in this way yield a fair estimate
of the upper limits of $\vsini$. The {\it micro-}turbulent velocity has
been adopted as $\vmicro=10\,\kms$. Even though this parameter is not
directly measurable (again, due to missing metallic lines), this value is
consistent with evidence from O/early B-stars in the considered parameter
range \citep{Villamariz00}.

The atmospheric parameters were determined from the hydrogen Balmer lines.
\Hg, \Hb, and \Ha, as well as the \HeI\ singlets \lam \lam 4387, 4922, the
\HeI\ triplets \lam \lam 4471, 4713 and the \HeII\ \lam \lam 4541 and 4686
lines (in case of \Htoe, \HeII\ \lam 4200 was available as well). In a first
step, we compared the observed profiles to synthetic ones from our grid of
OB star atmospheres \citep{Puls05}, leading to rough\footnote{Note that the
considerable differences in radii between normal OB-stars and CSPN (factors
of 5 to 10) deteriorate any estimates relying on the usual scaling relations
of wind diagnostics when using a grid of ``normal'' stars to investigate
central stars: the wind conditions do not depend on the wind-strength
parameter alone, but also on quantities with a different scaling behavior,
such as the density (see, e.g., \citealt{Puls05}).} estimates for $\Teff$,
$\lg$, the wind strength parameter $Q=\Mdot/(\vinf\,R)^{3/2}$ \citep{Puls96},
and the helium abundance $\yhe=N_{\rm He}/N_{\rm H}$.

To improve the solution, all atmospheric parameters (including the clumping
factor) have been fine-tuned to allow for the best possible fit, where
the specific radius of the model was chosen from the evolutionary tracks
(see Section~\ref{method1} and Figure~\ref{stellar_evo}), updated in
parallel with the progress of the line fits. The final model was re-run
with a radius adjusted to the $\lg$ corrected for rotational velocity
(see Section~\ref{method1} for details).

\begin{figure}
\resizebox{\hsize}{!}{\includegraphics{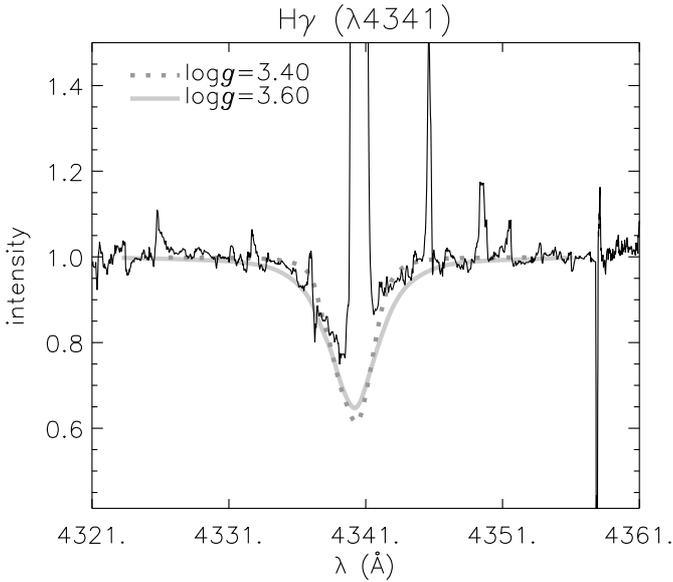}} \caption{\Mttt: \Hg\
at different gravities. $\Teff=39\,000$, $\yhe=0.1$, $\vsini=80$ and wind
strength as derived from our complete analysis (see Table~\ref{stapar}). Note
the difficulty in deciding on the best fit.} \label{Hg}
\end{figure}

At first, the line profiles of \Hg\ and \Hb\ were used to determine $\lg$
for the possible range in $\Teff$ as resulting from our crude estimates. The
effect of different gravities on \Hg\ is displayed in Figure~\ref{Hg}. Then,
the effective temperature was derived, from the {\it ratio} of \HeI\
to \HeII, whereas the absolute strengths of the helium lines define the
helium abundance, $\yhe$. Finally, the wind strength parameter, $\logQ$,
and the clumping factor $\fcl$ were determined, using all profiles, in
particular \Ha\ and \HeII\ 4686.

In case of strong mass-loss (well-developed \Ha), $\vinf$ could be
determined from the width of \Ha\ (we used $\vinf / \vesc \approx 2$ to $3$
as a starting point), whereas in cases of weaker $\Mdot$ the synthetic
profile reacts too weakly on variations of $\vinf$ to allow for such a
procedure. For these stars $\vinf$ was interpolated from the Table provided
by \citet{Kudritzki97}, who used terminal velocities for CSPN of roughly
similar parameters derived from UV resonance line profiles.

For all objects, we also checked the available silicon lines for
consistency. The corresponding model atom is the same as used and described
by \citet{Trundle04} in their analysis of SMC B supergiants.

\subsection{Wind clumping}
\label{clumping}

During our first analysis of the spectra, for two of our objects, \Mote\ and
\Mtot, it was impossible to fit \Ha\ and \HeII\ \lam 4686 at the same
mass-loss rate simultaneously (the predicted \HeII\ emission was by far too
large, compared to the observed absorption profiles, see Figure~\ref{m212},
dotted profiles).

Since at that time evidence had accumulated that WR and OB-star winds might
be clumped (cf. \citealt{Puls06} and references therein), and since it was
to be expected that clumping should affect \Ha\ and \HeII\ in a different
way for cooler objects, we investigated to what extent this process is
able to solve the discussed problem. The assumption of a clumped wind leads to
a satisfying solution.

Independent of this work, the same problem and solution was found in the
re-analysis of a different sample of CSPN, and a detailed discussion of the
involved physics can be found in the corresponding publication
\citep{Kudritzki06}. In brief, and as outlined above, small-scale clumping
preserves the profiles of $\rho^2$-dependent lines, as long as the {\it
apparent} mass-loss rate, $\Mdot \sqrt{\fcl}$, remains constant. Though for
almost the complete OB star range \Ha\ follows this scaling (since \HI\ is a
trace ion), \HeII\ becomes the dominant ion (or close to this) for late
O-type objects, and the corresponding opacities scale (almost) {\it linearly}
with density. In such a case then, the corresponding line profiles (in
particular, \HeII\ \lam 4686) do {\it not} react on wind-clumping, but depend
on the actual mass-loss rate, $\Mdot$, alone. Thus, a clumped wind with a
mass-loss rate lower than derived from unclumped models can explain that
\HeII\ \lam 4686 is still in absorption (because the actual mass-loss rate is
low), whereas the strong emission in \Ha\ is a consequence of clumping.
Figure~\ref{m212} illustrates this behavior, where the solid grey profiles
have been calculated with a mass-loss rate reduced by a factor of
$\sqrt{\fcl}$ (compared to the dotted solution), and a clumped wind with
$\fcl = 30$ has been assumed. Note that all other lines analyzed here react
only weakly to the presence/absence of clumping, either because they are
formed very close to the photosphere, or because they are also dependent on
$\rho^2$ (such as \HeI\ at cooler temperatures).

Two additional comments: (i) For hotter objects ($\Teff
\ga$~35{\ldots}38\,kK, depending on gravity), \HeIII\ becomes the dominant
ion, and {\it both} \Ha\ and \HeII\ \lam 4686 depend on the apparent
mass-loss rate. In such a case, the comparison of both lines does not
allow us to conclude on the degree of clumping. (ii) If the observed \Ha\
and \HeII\ \lam 4686 lines show a similar degree of emission at {\it
cooler} temperatures, this strongly suggests that the corresponding wind
is un\-clumped, i.e., $\fcl=1$.

\subsection{Error estimates for quantities derived from the spectra alone}

Thanks to the high quality of our spectra, fitting errors due to resolution
limitations or instrumental noise do not play a role in our analysis. The
major problem encountered here is the strong contamination of the hydrogen
and \HeI\ lines by nebular emission in their cores and beyond. In certain
cases (particularly in \Ha, e.g. \Mots\ and \Mote) the contamination is
so severe that the uncertainties in the derived parameters depend mainly
on our inability to identify the {\it stellar} profile unambiguously.

Apart from this principal problem, the second major source of errors is
due to our ``eye-fit'' procedure (contrasted to automated methods, e.g.,
\citealt{Mokiem05}), which in certain cases might not be able to find the
{\it global} optimum in parameter space. Because of the severe nebular
contamination, however, an automated method is very difficult to apply,
due to the problems of defining the ``true'' stellar spectrum required
for such a procedure.

In the following, we will summarize the typical errors resulting from our
{\it spectral} analysis. Errors of quantities related to other diagnostics
(masses, radii, distances and absolute mass-loss rates) will be discussed
in Section \ref{deduced}.

\subsubsection{Effective temperatures}

Depending on the quality of the helium line fits and the nebula
contamination, the typical formal\footnote{i.e., not accounting for any
errors intrinsic to the synthetic profiles.} errors in $\Teff$ are about
$\pm 1\,500$\,K. This uncertainty becomes larger in the region around $\Teff
= 32\,000{\ldots} 36\,000$\,K, if we account for the (known) problem of
a possible inconsistency between the predicted line-strengths of \HeI\
triplets and singlets (due to subtle line overlap effects regarding
the \HeI\ resonance line in the FUV, see \citealt{Najarro06}). In this
temperature range, the triplet lines are the more reliable ones (see also
\citealt{Mokiem05}), and are consequently used as a main indicator for
the determination of $\Teff$.

\begin{figure*} \resizebox{\hsize}{!}{\includegraphics{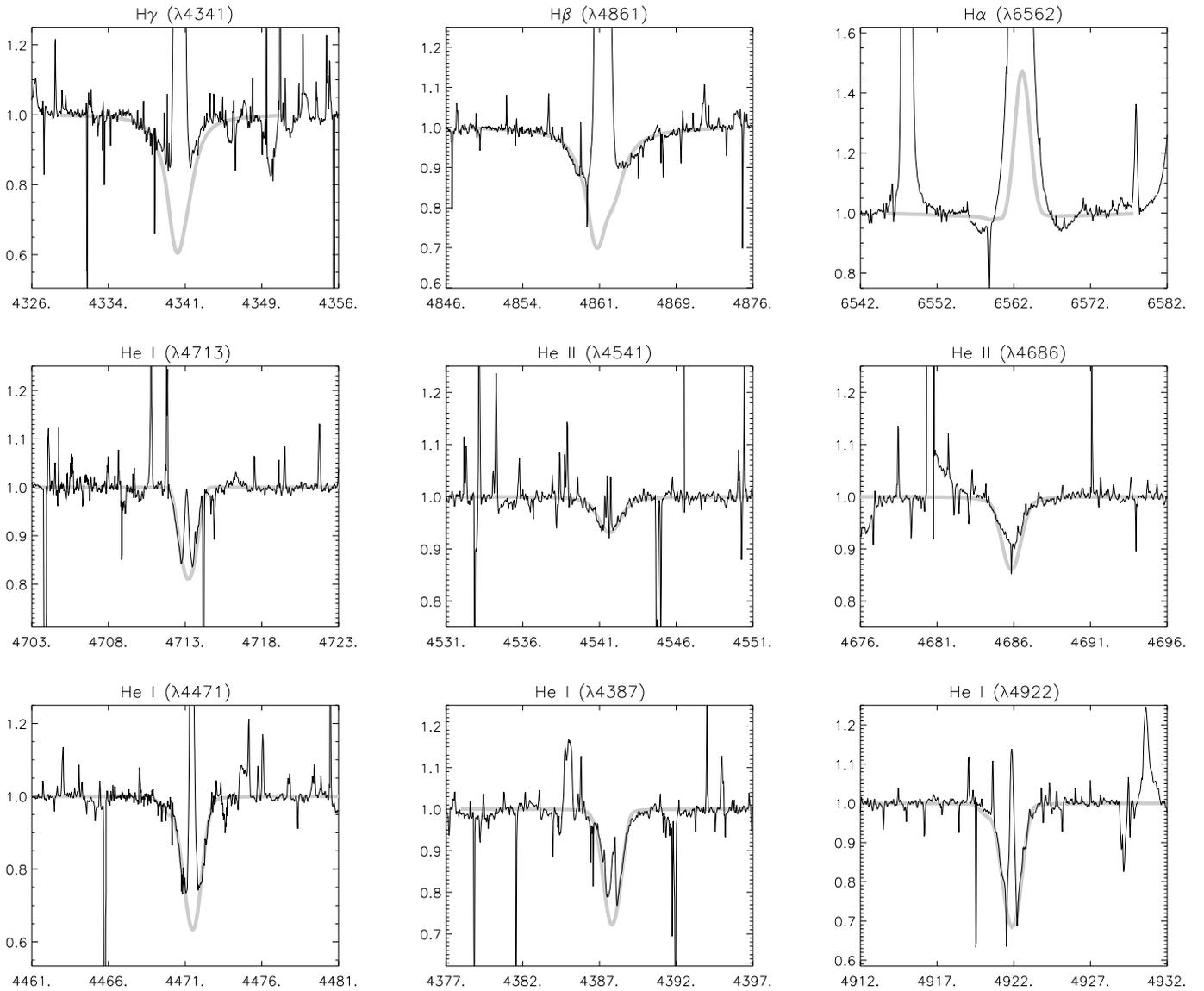}}
\caption{\Htts: H and \HeI/\HeII\ lines analyzed in this study,
together with the predictions from our best fitting model (grey).}
\label{h226} \end{figure*}

\begin{figure*} \resizebox{\hsize}{!}{\includegraphics{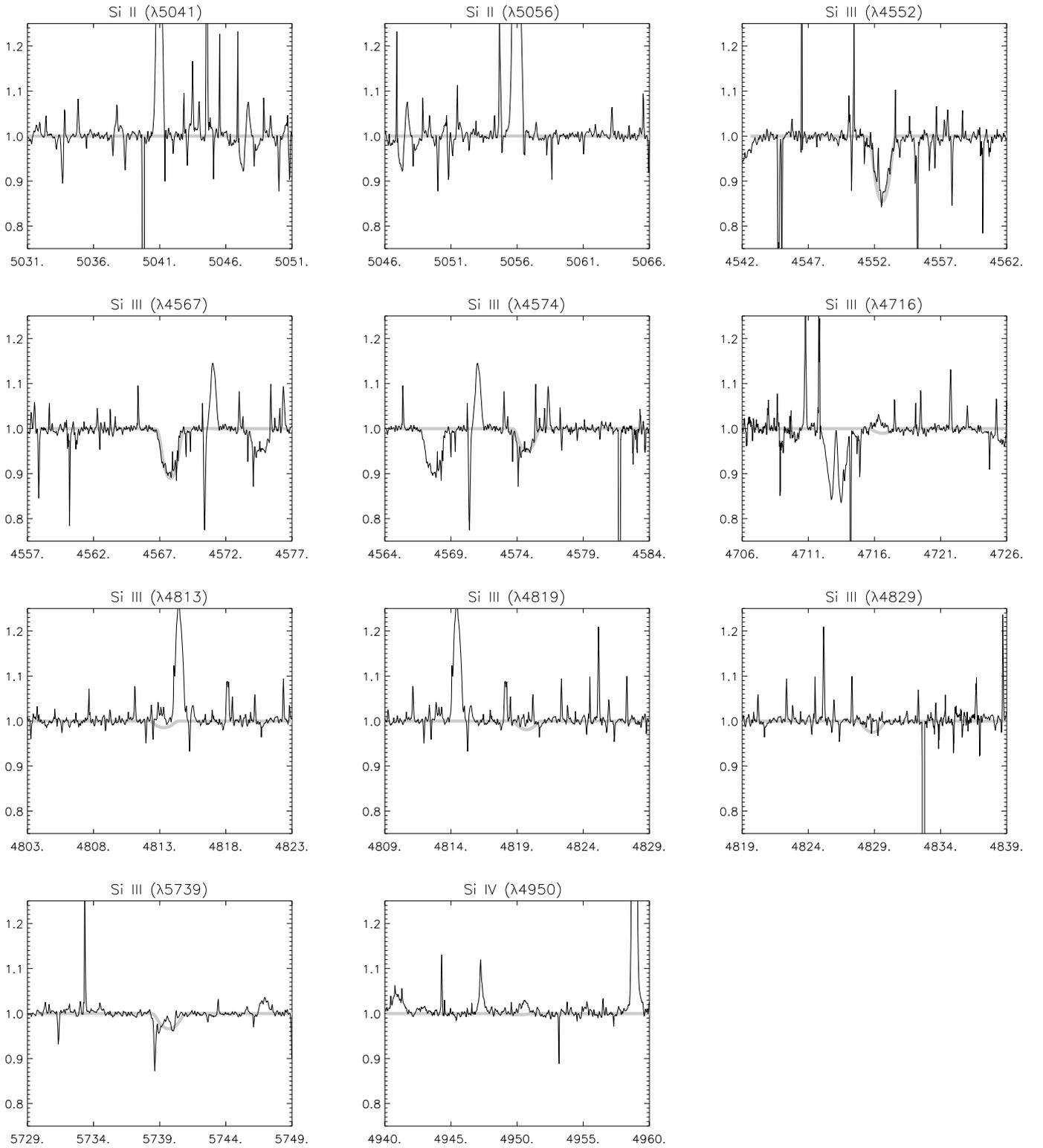}}
\caption{\Htts: important Si lines present in our data set. Over\-plotted are
the theoretical predictions for identical parameters as used in
Figure~\ref{h226} (solar Si abundance).} \label{h226Si} \end{figure*}

\begin{figure*} \resizebox{\hsize}{!}{\includegraphics{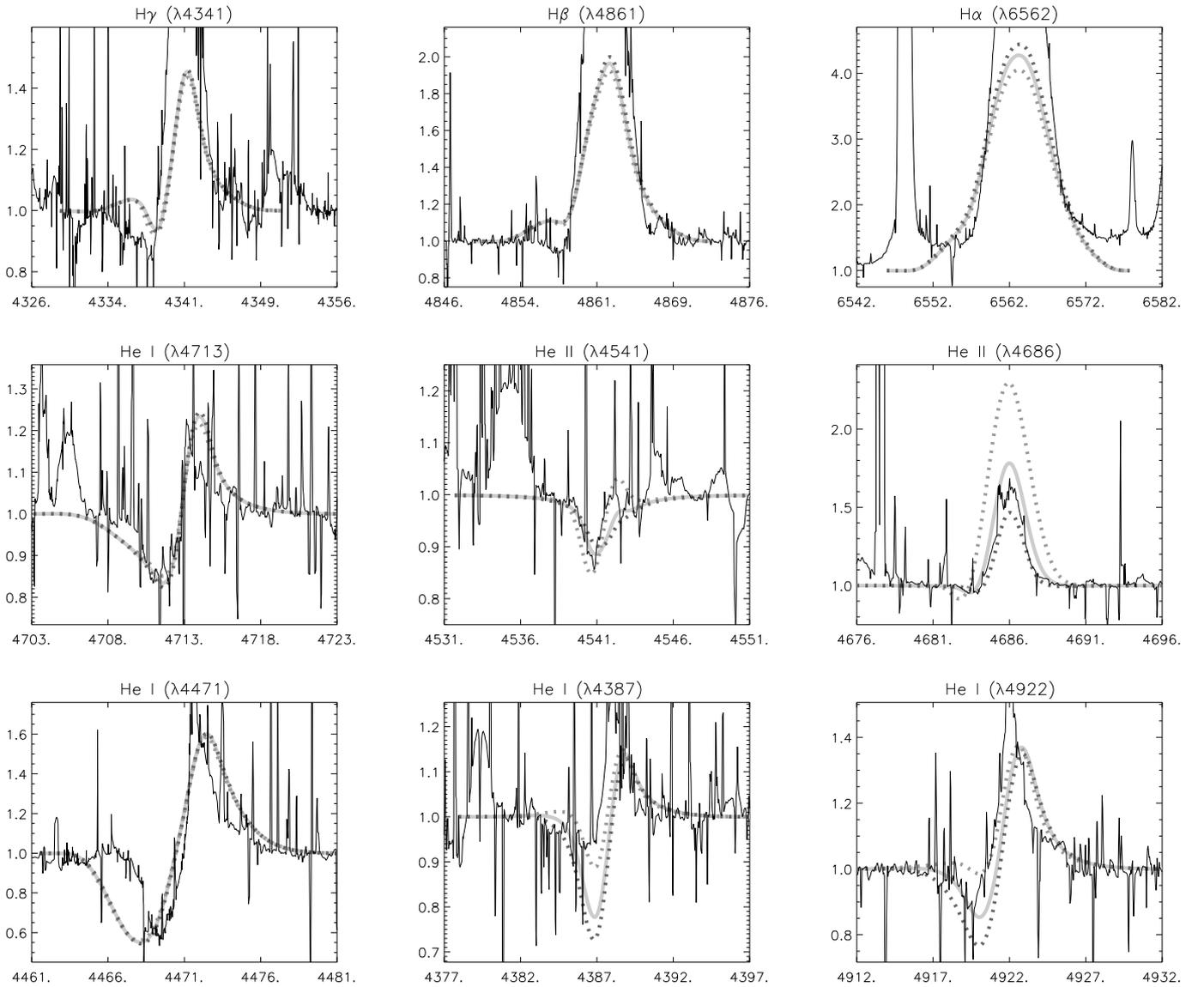}} \caption{As
Figure~\ref{h226}, but for \Mots. The dotted profiles refer to models with
different gravities, $\Delta \log g = \pm 0.3$ (stronger emission in \HeII
\lam 4686 for decreased gravity, see text).} \label{m137} \end{figure*}

\begin{figure*} \resizebox{\hsize}{!}{\includegraphics{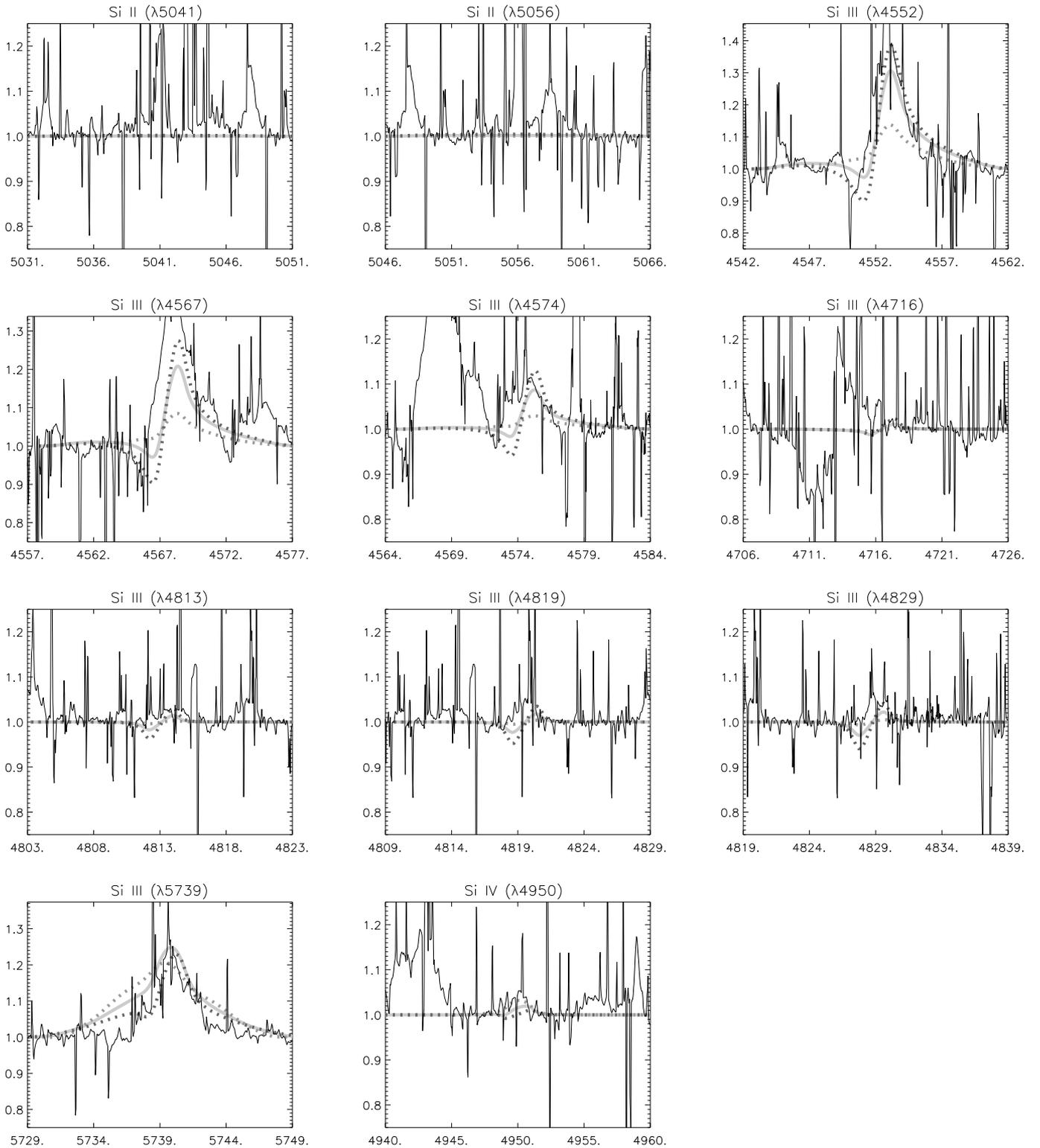}}
\caption{As Figure~\ref{h226Si}, but for \Mots. A silicon abundance of
roughly 30 times the solar one has to be adopted in order to reproduce the
observations.  Again, the dotted profiles refer to models with different
gravities, $\Delta \log g = \pm 0.3$.} \label{m137Si} \end{figure*}

\begin{figure*} \resizebox{\hsize}{!}{\includegraphics{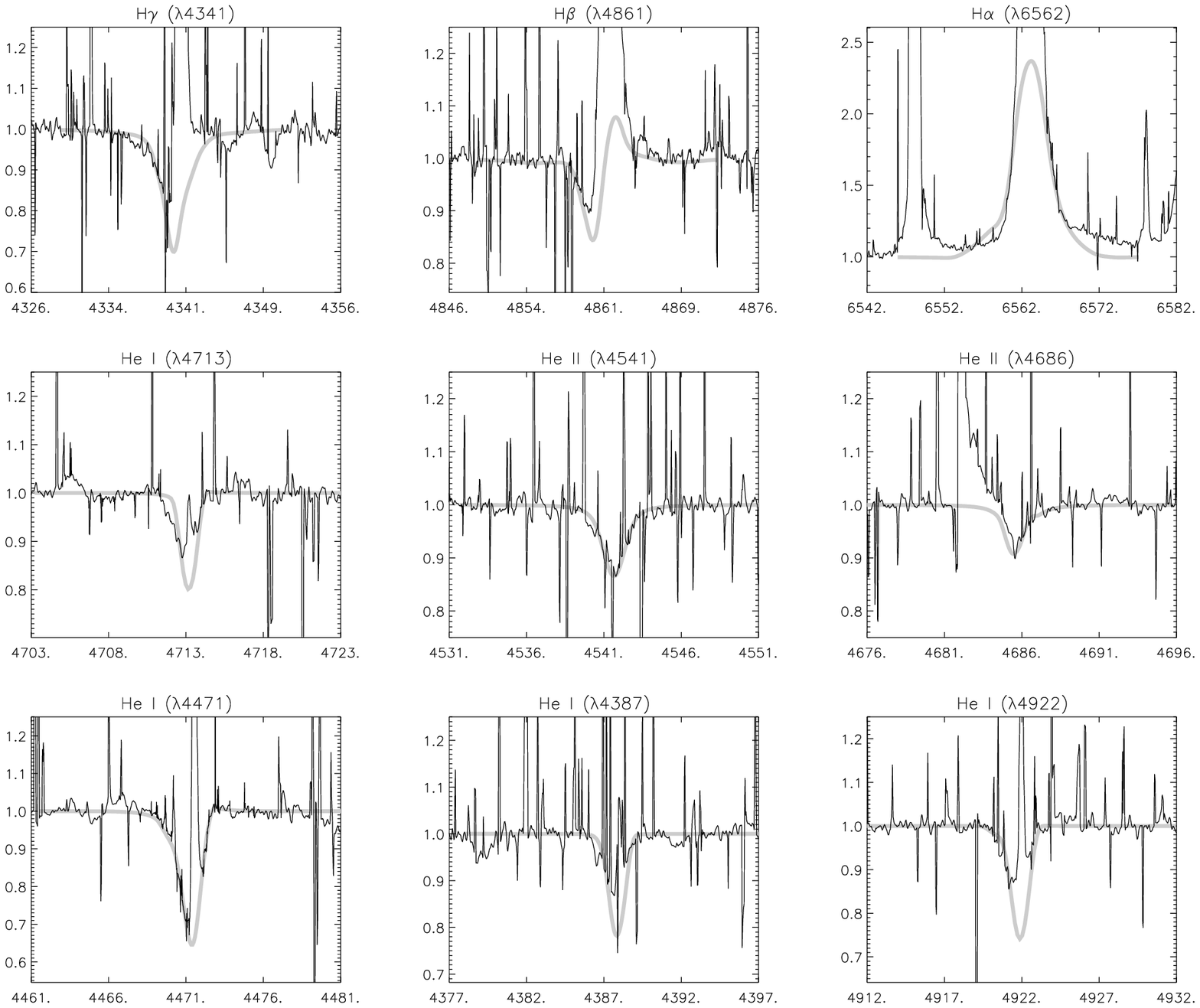}} \caption{As
Figure~\ref{h226}, but for \Mote.} \label{m138} \end{figure*}

\begin{figure*} \resizebox{\hsize}{!}{\includegraphics{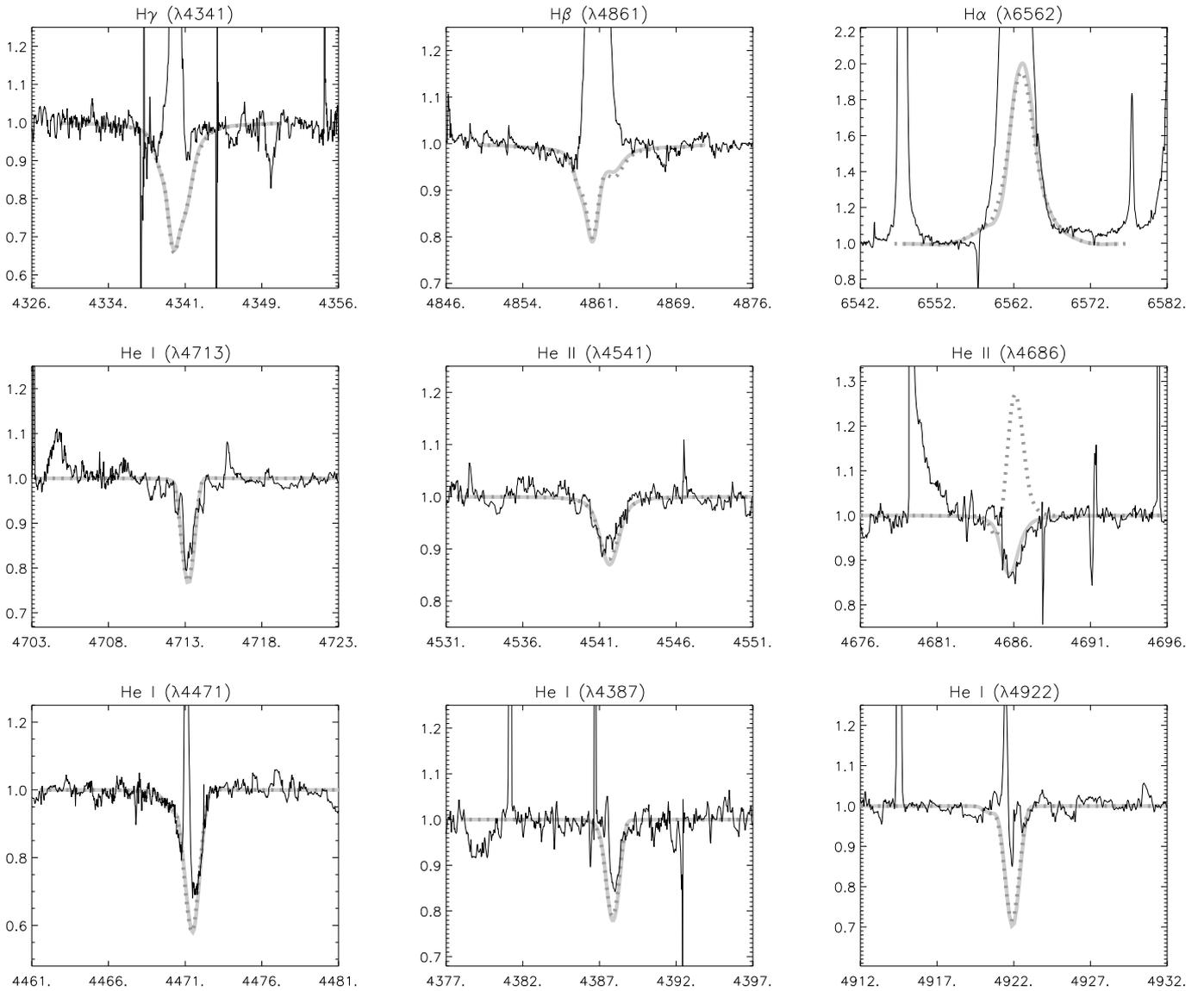}} \caption{As
Figure~\ref{h226}, but for \Mtot. The dotted profiles display the predictions
from an {\it unclumped} model at larger mass-loss rate. Note that the only
difference regards \HeII\ \lam 4686 (see Section~\ref{clumping}).}
\label{m212} \end{figure*}

\begin{figure*} \resizebox{\hsize}{!}{\includegraphics{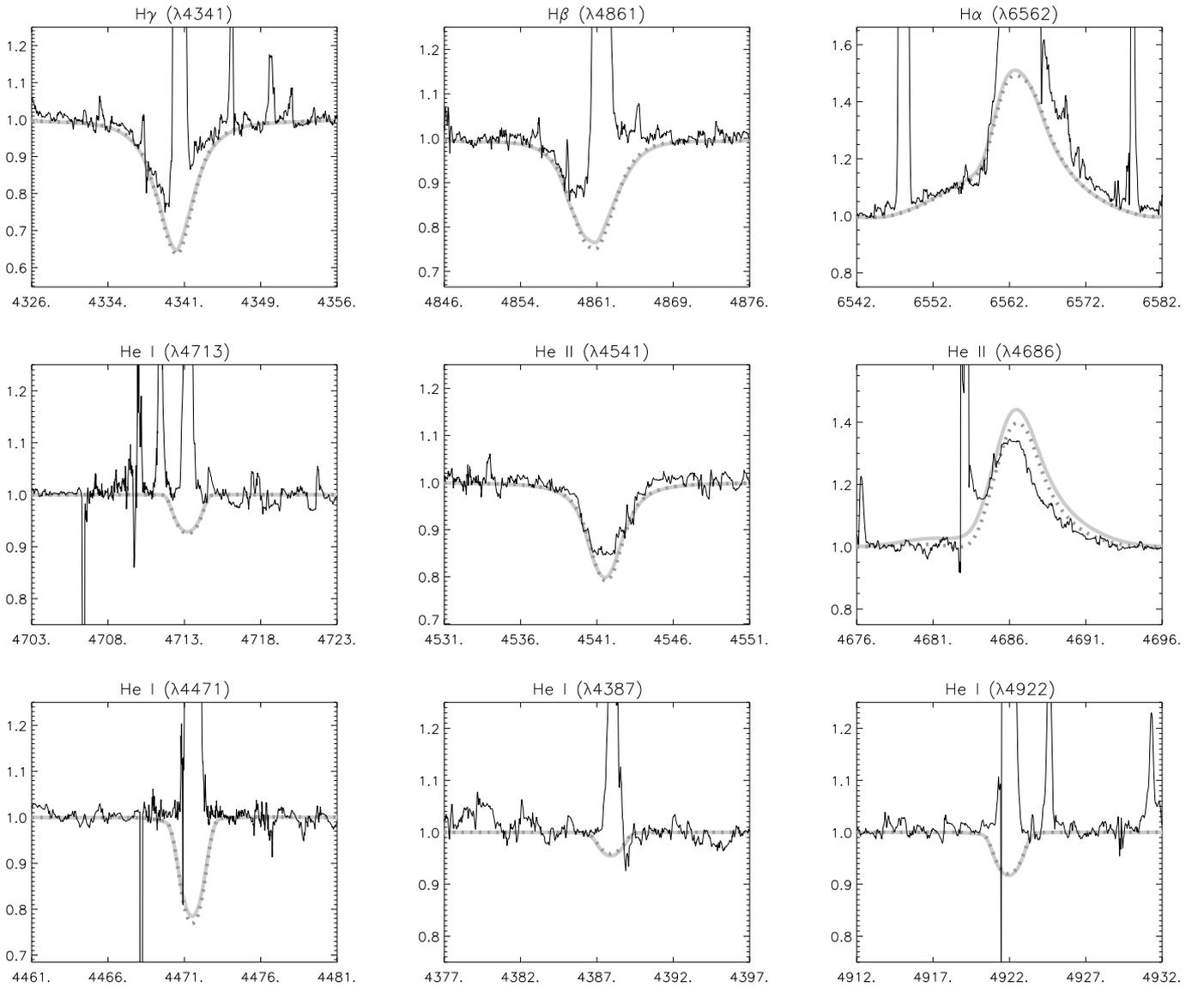}} \caption{As
Figure~\ref{h226}, but for \Mttt. The dotted profiles refer to a clumped
model with $\fcl=7$ (see text).} \label{m233} \end{figure*}

\begin{figure*} \resizebox{\hsize}{!}{\includegraphics{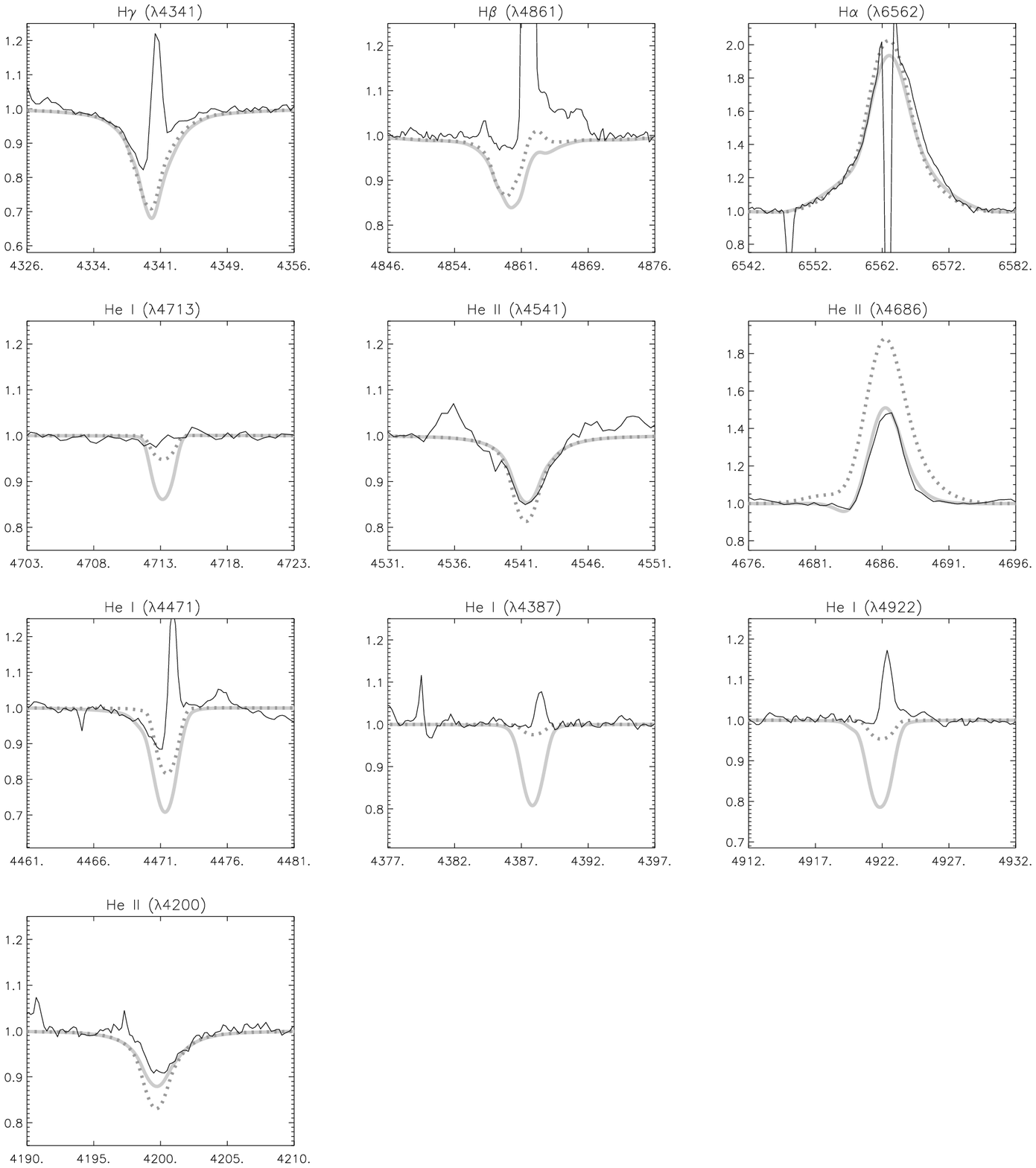}}
\caption{\Htoe, H and He line fits. Note the lower quality of the observed
spectra, compared to the other objects. Solid: low $\Teff$ solution, for an
unclumped model and a mass-loss rate required to fit \Ha\ and \HeII\ \lam
4686. Dotted: high $\Teff$ solution, again for an unclumped model designed to
match \Ha. (see text).} \label{h218} \end{figure*}

\subsubsection{Gravities}

Because of the intrinsic coupling of $\lg$ and $\Teff$ when deriving the
gravity from the Balmer line wings (typically, a change by 1000\,K
corresponds to change in $\lg$ by 0.1, in the same direction), the precision
of $\lg$ (which is of major importance in the present investigation) depends
strongly on the precision of $\Teff$ (from the He ionization balance) which
in turn requires a reasonable determination of the helium abundance, $\yhe$.
With respect to the available spectra and their nebular contamination, this
may not be warranted in all cases. Particularly for He 2-108, the
uncertainties were large enough to persuade us to provide two different
solutions for this object, at lower and higher $\Teff$ (for details, see
Section~\ref{he2108_hilow}).

In most other cases, we adopt, in accordance with similar investigations,
a typical error of $\Delta \lg = \pm 0.1$, since for larger differences the
profile shape changes significantly. If the star is close to the Eddington
limit, the effective gravity becomes smaller and smaller because of the
increased radiation pressure (not only from Thomson scattering, but also
from photospheric {\it line} pressure). The ``true'' gravity is given by
$g_{\rm grav} = g_{\rm eff} + g_{\rm rad}$ (excluding centrifugal forces
for the moment, but see Section~\ref{method1}), where the Balmer line
wings actually ``measure'' $g_{\rm eff}$ because of their dependence on the
electron density. As soon as $g_{\rm rad} \ga g_{\rm eff}$, any uncertainty
in the calculated radiation pressure (intrinsic to the atmospheric models
and strongly dependent on the completeness of the line list) results in a
larger error for $g_{\rm grav}$, which we adopt in these cases by $\Delta
\lg = 0.15$.

\subsubsection{Wind parameters}

As outlined above, terminal velocities were derived, where possible, from
the width of \Ha. Particularly, we found $\vinf = 500\pm100$\,km/s for
\Mots, and $\vinf = 800\pm160$\,km/s for \Mttt. Strictly speaking, both
values can be considered as lower limits only. For the other objects we
adopted $\vinf$ from interpolating between ``known'' terminal velocities
measured for different samples (cf. \citealt{Kudritzki97}), and allowed
for typical errors of 20\%.

The precision of the wind-strength parameter, $\logQ$, on the other
hand, depends on the validity of the derived velocity exponents, $\beta$
\citep{Puls96}. To account for this fact, we have fitted $\logQ$ for
different values of $\beta \pm \Delta \beta$, with $\Delta \beta = 0.1$. In
this case then, $\Delta \logQ$ is typically not larger than $\pm 0.1$
(see also \citealt{Markova04} and \citealt{Repolust04}).

\subsection{Comments on the individual objects}
\label{comments}

In the following, we compare our best fitting synthetic spectra with the
observed ones, for all H and He lines which have been analyzed. For two of
our objects, \Htts\ and \Mots, the Si line ``fits'' are displayed as well.
Whereas for all but one of our objects the Si lines matched nicely the
observations at solar abundance (with \Htts\ being a prototypical object in
this respect), we found a very large Si-abundance for \Mots, which is still
lacking any theoretical explanation.

Note that all of our objects display a rather strong \Ha\ emission,
immediately pointing to large wind-strengths (cf.\ Table~\ref{stapar}),
which indeed are a factor of about 5 to 10 larger than for typical
O and B supergiants. Further comments on this behavior are given in
Section~\ref{mdotdmom}.

All results from our analysis including corresponding errors have been
summarized in Table~\ref{stapar}.

\subsubsection{\Htts}

For this star, we obtained a convincing fit, displayed in Figures~\ref{h226}
and \ref{h226Si}. The observed and synthetic \Hg\ lines agree well, i.e.,
$\lg$ is well determined within the nominal error. The second diagnostics for
$\lg$, \Hb, would require a slightly larger value, but this would deteriorate
not only \Hg\ but also \mbox{\HeI\ \lam\lam 4387 and 4922} as well as \SiIII.

Compared to the other objects, \Ha\ shows the lowest emission, which
immediately points to a ``low'' mass-loss rate. Since \HeII\ \lam 4686 is
closely reproduced in parallel (no clumping required to keep this line in
absorption at the derived effective temperature of 28\,000\,K), and all other
lines fit almost perfectly (including silicon, where \SiIII\ is matched at
solar abundance, and \SiII\ and \SiIV\ are absent as predicted) we feel
rather confident about the stellar and wind parameters of our model.

\subsubsection{\Mots}
\label{mots}

Although not classified as a Wolf-Rayet star, this star shows well
developed P-Cygni profiles in \HeI\ (Figure~\ref{m137}) and \SiIII\
(Figure~\ref{m137Si}), immediately pointing to an extremely dense wind
(the densest in our sample).

To fit the \HeI\ lines, an enhanced helium abundance of at least $\yhe
= 1.0$ is required, where even larger values cannot be excluded due to
saturation effects. For lower helium content, on the other hand, a fit
to the \HeI\ lines would require a larger mass loss rate, leading to too
strong emission in \Ha, \Hb\ and \Hg.

Although there is some (stellar?) absorption visible in the blue wings of
\Hg\ and \Hb, we were not able to match these features, thus leaving the
gravity rather unconstrained: Models calculated with different $\lg$ (and
all other parameters kept constant) display only marginal differences in
the spectra, which is not surprising since for such large wind-densities all
lines are formed in the wind. However, if we changed not only $\lg$ but also
$\Rstar$ (and $\Mdot$ to conserve $Q$) according to the core mass-luminosity
relation, differences became visible, because not {\it all} quantities scale
exactly with $Q$ (cf. \citealt{Puls05}). The differences for $\Delta \lg
= \pm 0.3$ can be seen in Figs.~\ref{m137} and \ref{m137Si}. The largest
reaction is found for \HeII\ \lam 4686, where the emission becomes stronger
with lower $\lg$ and larger $\Rstar$. Note, however, that this effect can
be compensated by introducing a certain degree of clumping (see below).

As we were not able to derive $\lg$ spectroscopically, we {\it adopted}
$\lg \approx 3.4$ from distance considerations (assuming that the star is
located at $\approx$ 8\,kpc). Lower values for $\lg$ would place the star
further away, higher values to closer distances. Moreover, we adopt quite
a large error of $\Delta \lg = \pm 0.3$, thereby allowing for distances
in between 5.3\,kpc and 11.5\,kpc (see Table~\ref{stapar}). This seems
to be a reasonable range, comparing with our most distant object, \Mtot,
located at 11.4\,kpc according to our analysis.

The determination of $\Teff$ from the ratio of \HeI\ to \HeII\ provides
similar problems as the definition of the other parameters.  The adopted
value of 31\,000\,K is an upper limit, depending alone on our fit to \HeII\
\lam 4541. A lower limit on $\Teff$ can be derived from the absence of
stellar \SiII\ lines, which have been found to appear below $\Teff =
29\,000$\,K. To fit the \SiIII\ emission lines, an extremely large Si
abundance is required (28 times the solar abundance). According to stellar
evolution models the Si abundance should not change, even in H-deficient
stars \citep{Werner06}, and we have no explanation for this finding.
\cite{Leuenhagen98} claim solar Si abundance for similar objects but mention
overabundances (up to 40 times solar) for some [WCL] stars.

Both \Ha\ and \HeII\ \lam 4686 are consistent with an unclumped wind of
the quoted strength, although the complete \Ha\ line is hidden within the
strong nebular emission. Note that there is no indication of clumping at
the derived/adopted parameters of our final model, since the presence of
clumping would decrease the \HeII\ emission and require a higher $\Teff$
to recover the observations, which is inconsistent with the other lines.
Nevertheless, (moderate) clumping cannot be excluded, since a lower gravity
(compensating for the decreased emission) is still possible. Note also
that a clumped wind might be able to decrease the derived Si abundance.

\subsubsection{\Mote}

All parameters can be derived with a reliable precision within the
nominal errors. $\lg$\ is well defined from the wings of \Hg\ and \Hb,
and $\Teff$\ and (solar) $\yhe$\ are equally well constrained from \HeII\
\lam\lam 4541/4686 and \HeI\ \lam 4471. In contrast, however, we could not
obtain a satisfying fit for the \mbox{\HeI\ \lam\lam 4387/4922} singlets
(overestimated strengths, presumably due to the ``singlet problem''
outlined above), and we have discarded them from our analysis.

This object is the first where we needed a clumped wind, in order to fit \Ha\
and \HeII\ \lam 4686 simultaneously at the derived effective temperature.
Note that all other lines do not react on clumping, but are consistent with
the derived ``apparent'' wind-strength, $\Mdot \sqrt{\fcl}$. This value and
the individual quantities for $\Mdot$ and $\fcl \approx 20$ were determined
by a compromise to fit the wind-induced asymmetries seen in most lines, the
emission profile in \Ha\ and the absorption line profile of \HeII\ \lam 4686.

\subsubsection{\Mtot}

Both \Hg\ and \Hb\ are dominated by the strong nebular lines, and only the
extreme wings become visible. This leads to a higher uncertainty in $\lg$,
with an error of $\Delta \lg = +0.2/-0.1$.

As above, the synthetic \HeI\ singlets might appear as too strong, but in
this case the nebular contamination forbids a reliable estimate on whether
this is really the case. From the triplet lines and \HeII\ \lam\lam 4541/4686
then, we derive a well defined $\Teff$ and helium abundance $\yhe$, within
the nominal errors.  We prefer a somewhat reduced helium abundance, $\yhe =
0.08$, in order to fit the globally weak He-lines.

Again, a clumping factor of $\fcl \approx 30$ is required to keep \HeII\ \lam
4686 in absorption. Further details on this problem have been outlined in
Section~\ref{clumping}.

\subsubsection{\Mttt}

For this object, the quality of both the spectra and the fits is
satisfactory, and nominal errors apply for all parameters except for $\Teff$
and thus $\lg$ (see Figure \ref{m233}). Since most of the \HeI\ absorption
lines are very weak and thus dominated by nebular emission, the ratio of
\HeI\ to \HeII\ cannot be determined well enough to ascertain the effective
temperature precisely. We estimate the corresponding error as $\Delta \Teff =
\pm 2\,000$\,K.

Anyway, in this temperature range (roughly 39\,000\,K) \HeII\ is no longer a
dominant ion, and the corresponding opacities scale with $\rho^2$. According
to our arguments given in Section~\ref{clumping}, a unique determination
of the clumping factor is no longer possible, and we adopt $\fcl = 1$. A
comparison between such an unclumped and a clumped model is given in Figure
\ref{m233}: the dotted profiles refer to a clumped model with $\fcl=7$,
and the small difference in \HeII $\lambda$4686 result from the fact that
\HeIII\ recombines in the outmost wind ($v/\vinf \ge 0.8$).

\subsubsection{\Htoe}
\label{he2108_hilow}

This object has been previously analyzed in the literature, with two
differing values for $\Teff$ (and related quantities). A low temperature
solution with $\Teff$ = 35\,000\,K\,/\,34\,000\,K was deduced by Kudritzki et
al. (1997/2006), from the optical, whereas a high temperature solution with
$\Teff$ = 39\,000\,K was found by \citet{Pauldrach04}, on the basis of the
Fe{\sc iii}/{\sc iv} ionization equilibrium derived from the UV.

Given the low quality of the available spectra, none of these solutions
can be excluded. Three of the four available \HeI\ lines {\it appear} to
be weak, thus supporting the high temperature solution, whereas \HeI\ \lam
4471 seems to be much stronger, thus favoring the lower $\Teff$. Rather
than trying to find a compromise, we decided to show both solutions,
and to follow the consequences throughout the remainder of this paper.

Our low temperature solution (Figure~\ref{h218}, solid) has been optimized
for \HeI\ \lam 4471, and results in $\Teff$ = 34\,000\,K, identical to the
value derived by \cite{Kudritzki06}. In this case, both \HeI\ singlets
\lam \lam 4387/4922 and the \HeI\ \lam 4713 triplet are overestimated,
whereas for the high temperature solution (dotted) it is basically the
other way round.  Note that the singlets at $\Teff$ = 34\,000\,K might be
overestimated, due to the singlet problem. At least at $\Teff$ = 39\,000\,K,
however, this problem is no longer relevant, and the agreement between
observed and predicted singlets cannot be disputed.

The errors for our low temperature solution (from \HeI\ \lam 4471) have been
estimated as +1\,000/$-$2\,000\,K for $\Teff$, and a standard error of
$\Delta \lg = 0.1$. In this case then, the helium abundance must be slightly
enhanced, $\yhe = 0.15$.

From the simultaneous analysis of the Balmer lines and \HeII\ \lam 4686, we
found an interesting result on the clumping properties of the wind. In
contrast to the other objects, \Htoe\ seems to require larger clumping
factors in the inner wind, compared to the outer one. First, note that, at
$\Teff \ga 34\,000$\,K, there is no discrepancy between \Ha\ and \HeII\ \lam
4686 (in agreement with the discussion given in Section~\ref{clumping}). In
contrast, however, there is now an inconsistency between \Ha\ and \Hg/\Hb.

For the low temperature solution, we have ``normalized'' the (unclumped)
mass-loss rate to fit \Ha/\HeII\ \lam 4686, and obviously the cores of \Hg\
and \Hb\ are predicted as too deep. Thus, clumping must be larger in the
inner wind, to allow for a refilling of both lines due to increased emission.

For the high temperature solution, an unclumped model designed to fit \Ha\
predicts too much \HeII\ \lam 4686 emission (this inconsistency cannot be
cured by clumped models, because of the high temperature), and, again, too
little emission in \Hg\ and \Hb.

Note that \Htoe\ is the only object in our sample which indicates such a
stratified clumping. Because in the present analysis we are considering
only depth independent clumping (see Section~\ref{modelatm}), we provide
(both for the high and the low temperature solution) a clumping factor of
unity in Table~\ref{stapar}, consistent with the wind-lines, but are aware
of the fact that the real mass-loss rate(s) might be smaller than indicated.

\section{Deduced Parameters: Masses, radii, distances and wind-momentum rates}
\label{deduced}

\begin{table*}[t]
\caption{Results of our analysis for the complete sample, using either
``Method~1'' or ``Method~2'' (see text). $\log g_{\rm spec}$ refers to the
gravity as derived from spectroscopy, and $\log g_{\rm true}$ is an estimate
(upper limit) for the true gravity, corrected for centrifugal acceleration
and used to calculate related physical properties. Wind strength parameter
$\logQ$ calculated with $\Mdot$ in units $\Msun$/yr, $\vinf$ in km/s and
$\Rstar$ in $\Rsun$. Modified wind-momentum rate, $\Dmom$, calculated in cgs.
The different distances resulting from ``Method~1'' refer to the different
extinctions as given at the top of the table. All other parameters which are
dependent on extinction refer to the ``mean'' value, $A_{V \,{\rm mean}}$
(see text).} \begin{center}
\renewcommand{\arraystretch} {1.4}
\begin{tabular*}{\textwidth}{l@{\extracolsep\fill}|ccccc|cc}

\hline \hline

Object name & He 2$-$260 & M 1$-$37$^{***}$ & M 1$-$38 & M 2$-$12 & M 2$-$33 & He 2$-$108 & He 2$-$108 \\
PN G & $008.2+06.8$ & $002.6-03.4$ & $002.4-03.7$ & $359.8+05.6$ & $002.0-06.2$ & $316.1+08.4$ & $316.1+08.4$ \\

\hline

$ V $ \hspace{\fill}(mag)~ & $ 14.27 \pm 0.10 $ & $ 14.99 \pm 0.25 $ & $ 14.45 \pm 0.25 $ & $ 14.74 \pm 0.10 $ & $ 14.40 \pm 0.50 $ & $ 12.63 \pm 0.10 $ & $ 12.63 \pm 0.10 $ \\
$ A_{V ~{\rm optical}}$ ~ \hspace{\fill}(mag)~ & $  1.95 $ & $  2.49 $ & $  2.62 $ & $  2.55 $ & $  1.06 $ & $  1.30 $ & $  1.30 $ \\
$ A_{V ~{\rm radio}}$ ~ \hspace{\fill}(mag)~ & $  1.57 $ & $  1.57 $ & $  1.81 $ & $  1.81 $ & $  0.42 $ & $  0.94 $ & $  0.94 $ \\
$ A_{V ~{\rm mean}} \,^{*} $ \hspace{\fill}(mag)~ & $  1.76 \pm  0.19 $ & $  2.03 \pm  0.46 $ & $  2.21 \pm  0.41 $ & $  2.18 \pm  0.37 $ & $  0.74 \pm  0.32 $ & $  1.12 \pm  0.18 $ & $  1.12 \pm  0.18 $ \\

\hline

$ v_{\rm rad} $ \hspace{\fill}(km/s)~ & $   37 \pm 10 $ & $  235 \pm 10 $ & $  -75 \pm 10 $ & $  110 \pm 10 $ & $ -127 \pm 10 $ & $  -35 \pm 10 $ & $  -35 \pm 10 $ \\
$ v\,\sin i \,^{**} $ \hspace{\fill}(km/s)~ & $ <  55 \pm 10 $ & $ <  70 \pm 10 $ & $ <  50 \pm 10 $ & $ <  60 \pm 10 $ & $ <  80 \pm 10 $ & $ <  80 \pm 10 $ & $ <  80 \pm 10 $ \\

\hline

$ T_{\rm\kern-.15em ef\kern-.05em f} $ \hspace{\fill}(K)~ & $  28000~^{+1500}_{-1500} $ & $  31000~^{+1500}_{-2000} $ & $  31000~^{+1500}_{-1500} $ & $  30500~^{+1500}_{-1500} $ & $  39000~^{+2000}_{-2000} $ & $  34000~^{+1000}_{-2000} $ & $  39000~^{+1500}_{-1500} $ \\
$ \log g_{\rm spec} $ & $ 2.90~^{+0.10}_{-0.10} $ & $ 3.40~^{+0.30}_{-0.30} $ & $ 3.05~^{+0.10}_{-0.10} $ & $ 3.00~^{+0.20}_{-0.10} $ & $ 3.60~^{+0.15}_{-0.15} $ & $ 3.50~^{+0.10}_{-0.10} $ & $ 3.70~^{+0.10}_{-0.10} $ \\
$ \log g_{\rm true} $ & $ 2.95~^{+0.09}_{-0.09} $ & $ 3.45~^{+0.27}_{-0.27} $ & $ 3.08~^{+0.09}_{-0.09} $ & $ 3.05~^{+0.18}_{-0.09} $ & $ 3.65~^{+0.14}_{-0.14} $ & $ 3.56~^{+0.09}_{-0.09} $ & $ 3.74~^{+0.09}_{-0.09} $ \\
$ Y_{\rm He} $ \hspace{\fill}($= N_{\rm He}/N_{\rm H}$)~ & $ 0.10 \pm 0.02 $ & $ 1.00 \pm 0.50 $ & $ 0.10 \pm 0.02 $ & $ 0.08 \pm 0.02 $ & $ 0.10 \pm 0.02 $ & $ 0.15 \pm 0.02 $ & $ 0.10 \pm 0.02 $ \\

\hline

$ \log Q $ & $ -12.3 \pm 0.1 $ & $ -10.9 \pm 0.1 $ & $ -12.3 \pm 0.1 $ & $ -12.6 \pm 0.1 $ & $ -11.7 \pm 0.1 $ & $ -11.7 \pm 0.1 $ & $ -11.7 \pm 0.1 $ \\
$ f_{\rm cl} $ & $  1 $ & $  1 $ & $ 20 $ & $ 30 $ & $  1 $ & $  1 $ & $  1 $ \\
$ \beta $ & $ 2.5 \pm 0.1 $ & $ 3.0 \pm 0.5 $ & $ 1.5 \pm 0.1 $ & $ 1.5 \pm 0.1 $ & $ 1.0 \pm 0.1 $ & $ 1.2 \pm 0.1 $ & $ 1.5 \pm 0.1 $ \\
$ v_{\rm \infty} $ \hspace{\fill}(km/s)~ & $  450 \pm   90 $ & $  600 \pm  120 $ & $  450 \pm   90 $ & $  450 \pm   90 $ & $  800 \pm  160 $ & $  700 \pm   70 $ & $  700 \pm   70 $ \\

\hline
\multicolumn{8}{c}{Method 1\,$^*$}\\
\hline

$ M_{\rm evol} $ \hspace{\fill}($M_\odot$)~ & $ 0.71~^{+0.13}_{-0.06} $ & $ 0.58~^{+0.09}_{-0.06} $ & $ 0.77~^{+0.11}_{-0.10} $ & $ 0.78~^{+0.10}_{-0.16} $ & $ 0.64~^{+0.09}_{-0.05} $ & $ 0.59~^{+0.02}_{-0.02} $ & $ 0.60~^{+0.03}_{-0.02} $ \\
$ R_{\rm evol} $ \hspace{\fill}($R_\odot$)~ & $  4.69~^{+ 0.68}_{- 0.51} $ & $  2.38~^{+ 0.89}_{- 0.64} $ & $  4.19~^{+ 0.56}_{- 0.50} $ & $  4.38~^{+ 0.57}_{- 0.92} $ & $  1.99~^{+ 0.37}_{- 0.30} $ & $  2.13~^{+ 0.24}_{- 0.22} $ & $  1.73~^{+ 0.20}_{- 0.18} $ \\
$ \log \, (L/L_\odot)$ & $  4.09~^{+ 0.15}_{- 0.14} $ & $  3.68~^{+ 0.29}_{- 0.30} $ & $  4.17~^{+ 0.14}_{- 0.14} $ & $  4.18~^{+ 0.13}_{- 0.22} $ & $  3.92~^{+ 0.17}_{- 0.17} $ & $  3.74~^{+ 0.10}_{- 0.14} $ & $  3.80~^{+ 0.11}_{- 0.12} $ \\
$ \dot M $ \hspace{\fill}($M_\odot$/yr)~ & $  0.45 \times 10^{-7} $ & $  7.63 \times 10^{-7} $ & $  0.40 \times 10^{-7} $ & $  0.24 \times 10^{-7} $ & $  1.13 \times 10^{-7} $ & $  1.07 \times 10^{-7} $ & $  0.92 \times 10^{-7} $ \\
$ \log \, D_{\rm mom}$ & $ 26.44~^{+ 0.25}_{- 0.24} $ & $ 27.65~^{+ 0.35}_{- 0.35} $ & $ 26.37~^{+ 0.25}_{- 0.25} $ & $ 26.15~^{+ 0.25}_{- 0.30} $ & $ 26.91~^{+ 0.27}_{- 0.26} $ & $ 26.84~^{+ 0.17}_{- 0.17} $ & $ 26.73~^{+ 0.17}_{- 0.17} $ \\

\hline

$ d_{\rm optical} $ \hspace{\fill}(kpc)~ & $  10.3~^{+  3.8}_{-  2.7} $ & $   6.3~^{+  3.5}_{-  2.2} $ & $   7.9~^{+  3.0}_{-  2.2} $ & $   9.6~^{+  3.4}_{-  3.0} $ & $   8.8~^{+  4.3}_{-  2.9} $ & $   3.4~^{+  1.2}_{-  0.9} $ & $   3.0~^{+  1.1}_{-  0.8} $ \\
$ d_{\rm radio} $ \hspace{\fill}(kpc)~ & $  12.2~^{+  4.5}_{-  3.2} $ & $   9.6~^{+  5.3}_{-  3.4} $ & $  11.5~^{+  4.4}_{-  3.2} $ & $  13.6~^{+  4.9}_{-  4.2} $ & $  11.9~^{+  5.8}_{-  3.9} $ & $   4.0~^{+  1.4}_{-  1.0} $ & $   3.6~^{+  1.3}_{-  0.9} $ \\
$ d_{\rm mean} $ \hspace{\fill}(kpc)~ & $  11.2~^{+  2.0}_{-  1.6} $ & $   7.8~^{+  3.8}_{-  2.6} $ & $   9.6~^{+  2.8}_{-  2.2} $ & $  11.5~^{+  2.8}_{-  2.9} $ & $  10.3~^{+  3.9}_{-  2.8} $ & $   3.7~^{+  0.6}_{-  0.5} $ & $   3.3~^{+  0.5}_{-  0.4} $ \\

\hline
\multicolumn{8}{c}{Method 2\,$^*$}\\
\hline

$ d_{\rm adopted} $ \hspace{\fill}(kpc)~ & $   8.0~^{+  2.0}_{-  2.0} $ & $   8.0~^{+  2.0}_{-  2.0} $ & $   8.0~^{+  2.0}_{-  2.0} $ & $   8.0~^{+  2.0}_{-  2.0} $ & $   8.0~^{+  2.0}_{-  2.0} $ & $   4.6~^{+  1.1}_{-  1.1} $ & $   4.6~^{+  1.1}_{-  1.1} $ \\
$ R_{\rm dist} $ \hspace{\fill}($R_\odot$)~ & $  3.36~^{+ 1.06}_{- 0.81} $ & $  2.44~^{+ 1.03}_{- 0.72} $ & $  3.51~^{+ 1.41}_{- 1.01} $ & $  3.06~^{+ 1.12}_{- 0.82} $ & $  1.55~^{+ 0.71}_{- 0.49} $ & $  2.64~^{+ 0.83}_{- 0.63} $ & $  2.41~^{+ 0.76}_{- 0.58} $ \\
$ M_{\rm dist} $ \hspace{\fill}($M_\odot$)~ & $ 0.38~^{+0.30}_{-0.17} $ & $ 0.61~^{+0.95}_{-0.37} $ & $ 0.55~^{+0.57}_{-0.28} $ & $ 0.40~^{+0.44}_{-0.19} $ & $ 0.40~^{+0.51}_{-0.22} $ & $ 0.89~^{+0.72}_{-0.40} $ & $ 1.14~^{+0.91}_{-0.51} $ \\
$ \log \, (L/L_\odot)$ & $  3.80~^{+ 0.26}_{- 0.26} $ & $  3.70~^{+ 0.32}_{- 0.33} $ & $  4.01~^{+ 0.31}_{- 0.31} $ & $  3.86~^{+ 0.28}_{- 0.29} $ & $  3.70~^{+ 0.34}_{- 0.34} $ & $  3.93~^{+ 0.24}_{- 0.26} $ & $  4.08~^{+ 0.25}_{- 0.25} $ \\
$ \log \, D_{\rm mom}$ & $ 26.15~^{+ 0.33}_{- 0.33} $ & $ 27.67~^{+ 0.38}_{- 0.38} $ & $ 26.21~^{+ 0.37}_{- 0.37} $ & $ 25.83~^{+ 0.35}_{- 0.35} $ & $ 26.69~^{+ 0.39}_{- 0.39} $ & $ 27.03~^{+ 0.28}_{- 0.28} $ & $ 27.02~^{+ 0.28}_{- 0.28} $ \\

\hline \hline

\multicolumn{8}{l}{\small
$^{*}$ see text
~ ~
$^{**}$ upper limit, since $v_{\rm macro}$ not measurable
~ ~
$^{***}$ $\log g_{\rm spec}$ adopted, see text
\hspace{\fill}}\\

\end{tabular*}
\end{center} \label{stapar} \end{table*}

\subsection{Two different methods}
Having presented those results that can be derived from spectroscopy alone,
we will concentrate now on the ``remaining'' parameters that are not {\it
directly} deducible from our analysis, since they depend on the stellar
radius, $\Rstar$. Two different, independent approaches will be applied
to obtain these parameters: ``Method~1'' uses evolutionary tracks for
central stars to provide masses, which, together with $\log g$, result in
stellar radii and finally {\it spectroscopic} distances. ``Method~2'',
on the other hand, uses {\it adopted} (bulge-)distances to calculate
radii and depending quantities. We will give a detailed error analysis and
compare the results and inherent uncertainties of both methods. This will
in particular clarify (at least for our sample of objects) the question
raised in Section~\ref{intro}, namely in how far the {\it combination}
of post-AGB evolution theory and state-of-the-art model atmospheres can
successfully predict distances from high-resolution spectroscopy.

So far, the spectroscopically determined gravities (from the Balmer-line
wings, denoted by $\log g_{\rm spec}$ in Table \ref{stapar}) are effective
values, since centrifugal forces lead to an outward acceleration of the
atmospheric layers, thus reducing the ``true'' gravity, $\log g_{\rm true}$.
In order to obtain this quantity (which has to be used in our subsequent
analysis), we have to correct $\log g_{\rm spec}$ by adding the centrifugal
acceleration. This has been accomplished in the ``usual'' way by means of
the derived projected rotational velocities, $v \sin i$, and the average
value of $\sin i$ (for details, see Appendix in \citealt{Repolust04} and
references therein). As already outlined in Section~\ref{specan}, due to
missing metallic lines the macro-turbulent velocity, $v_{\rm macro}$, was not
measurable for our sample, which means that the quoted values of $v \sin i$
and thus the true gravities have to be considered as upper limits. Note,
however, that despite the rather low rotational speeds the differences
between $\log g_{\rm spec}$ and $\log g_{\rm true}$ are quite significant
(of the order of 0.05 dex), which is due to the small radii of our objects
(centrifugal acceleration $\propto v_{\rm rot}^2/\Rstar$).

\subsubsection{Method~1}
\label{method1}

\begin{figure}
\resizebox{\hsize}{!}{\includegraphics{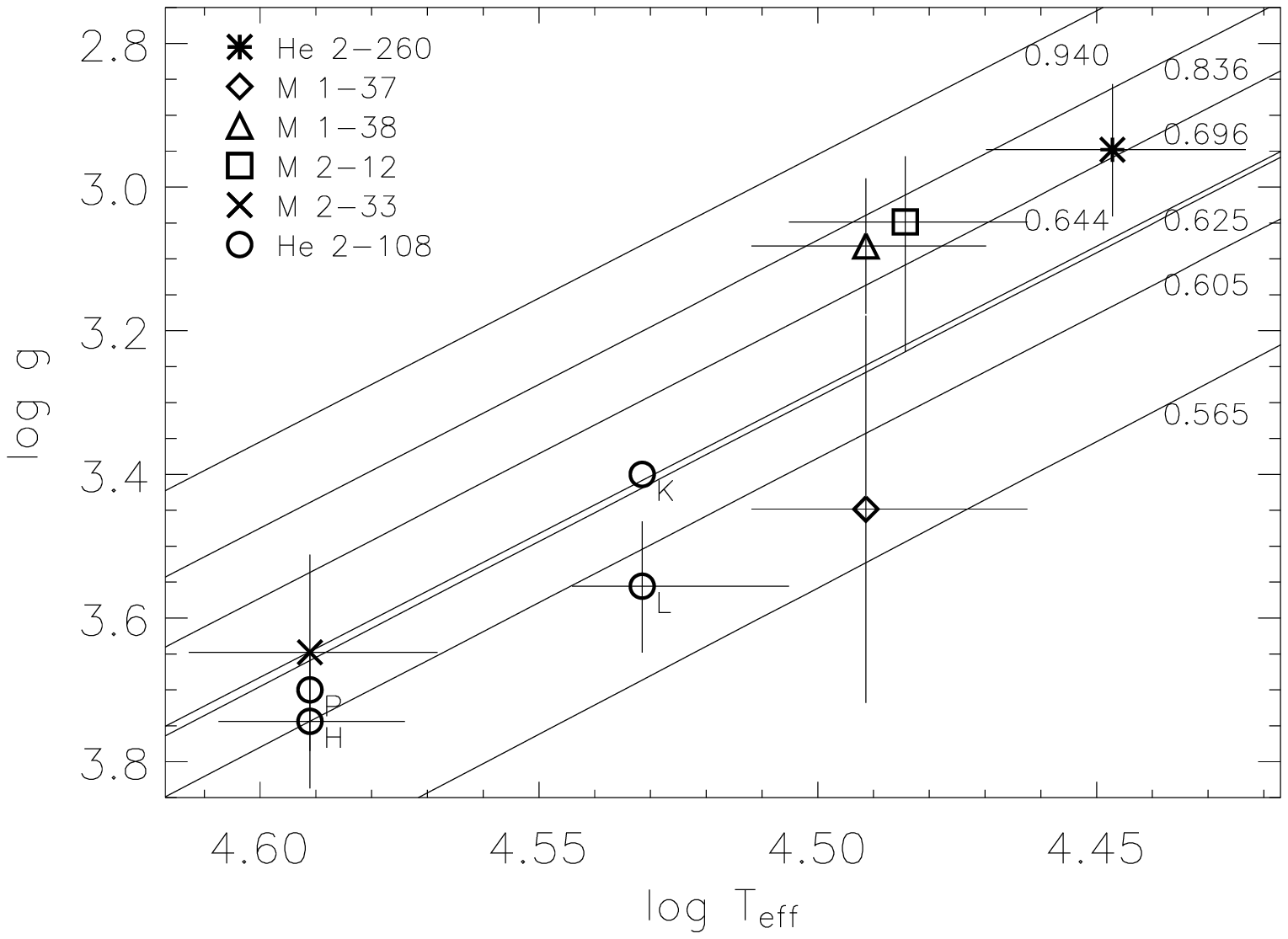}}
\caption{The positions of the CSPN in the $\log \Teff-\lg$~diagram,
with error bars according to the derived uncertainties, together with
theoretical post-AGB evolutionary tracks for stars of five different masses
(from \citealt{Bloecker95} and \citealt{Schoenberner89}, masses in units
of $\Msun$).  $L$, $H$, $K$ and $P$ denote various results for \Htoe:
our {\it l}ow and {\it h}igh temperature solution, the analysis by {\it
K}udritzki et al. (2006) and by {\it P}auldrach et al. (2004). Remember
that the gravity of \Mots\ has been {\it adopted}.} \label{stellar_evo}
\end{figure}

Here we apply the method to derive spectroscopic distances as described by
\cite{Mendez88}. We use the evolutionary tracks in the $\log \Teff$-$\lg$
diagram\footnote{Actually, those tracks are tabulated as a function of
luminosity and effective temperature, which can be easily converted into
the required $\lg$ vs. $\Teff$ diagram due to given masses.} for CSPN from
\cite{Bloecker95} and \cite{Schoenberner89} (Figure~\ref{stellar_evo}), which
defines the stellar mass, $\Mstar$, by interpolating the spectroscopically
determined $\Teff$ and $\log g_{\rm true}$ values between the evolutionary
tracks of different masses. This method is feasible for CSPN that are in
the phase of almost constant luminosity, because the evolutionary tracks
for different masses are unique in this regime.

Since the centrifugal correction term involves the stellar radius, an
iterative scheme is required: starting with some guess for $\Rstar$, we
obtain $\Mstar$, a new radius via $\log g_{\rm true}$, which is updated
subsequently and a new mass derived. The iteration process is continued
until mass and radius have converged. In Table~\ref{stapar}, we provide
these converged quantities and the corresponding true gravities. Note that
the masses and radii derived in this way depend on the validity of the
tracks (and on the accuracy of the spectroscopic analysis), but $\it not$
on the extinction.

\subsubsection{Method~2}

Here we exploit the fact that the observed objects are supposed to
be located in the Galactic bulge, following our selection criteria
(Section~\ref{selection}). We adopt the well-defined distance to the
Galactic center, $R_0$, of roughly 8\,kpc (\mbox{7.94 $\pm$ 0.42\,kpc},
\citealt{Eisenhauer03}) for all our objects, except for \Htoe.

For the latter one, we adopt a distance of 4.6\,kpc, which is consistent with
the value provided by \citet[ and also \citealt{Kudritzki06}]{Mendez92},
5.8\,{\ldots}6\,kpc, if we account for ``our'' extinction ($A_{V\,{\rm
mean}}$, see below) and visual magnitude $V$, together with their stellar
parameters. Admittedly, this distance is far from being as well determined
as $R_0$, and moreover it has been determined by using ``Method~1''. In
so far, the corresponding results should be considered as a comparison
rather than a test.

To check the error propagation of ``Method~2'' and to compare it with
``Method~1'', in all cases we adopt a typical error of 25\% in the
distance, which accounts for the depth of the Galactic bulge and follows
the considerations by \cite{Mendez88}.

From the distances, the visual magnitude, $V$ (cf. Table~\ref{stapar}, all
from \citealt{Tylenda91}) and the extinction, $A_V$ (see below), we find
the absolute magnitude, $M_V$, which together with the theoretical V-band
fluxes (convolved with the corresponding filter transmission function)
allows us to derive the stellar radius, $\Rstar$, following \citet[ see
also Eqs. (1,2) in \citealt{Repolust04}]{Kudritzki80}. The errors inherent
to all related quantities (mass, luminosity etc.) are thus dominated by
the uncertainty in distance and extinction, except for \Mttt, where the
large uncertainty in $V$ becomes decisive.

\subsubsection{Extinction}
\label{extinction}

As outlined in the introduction, \cite{Stasinska92} concluded, from a
comparison of Balmer decrement extinctions versus radio-H$\beta$ extinctions,
that for most distant PN the corresponding line of sight extinction law
should be significantly different from the standard law (for the diffuse
ISM). They argued that the ratio $R_V$ of total to selective extinction
(in the V-band) must be lower than the canonical value of 3.1, of the
order of $R_V$ = 2.7. Similar conclusions were reached more recently by
\citet{Udalski03} and \citet{Ruffle04}.

Using previous results by \citet{Seaton79}, the {\it standard} visual
extinction can be related to the logarithmic extinction at \Hb\ determined
from the Balmer decrement, $C_{\rm opt}$, via
\begin{equation}
A_{V \,{\rm optical}}=2.16 \, C_{\rm opt} \qquad (\mbox{``standard''
reddening}).
\end{equation}
The measurements of $C_{\rm rad}$ (the logarithmic extinction at \Hb\
determined from the radio fluxes) for a large sample of PN by Stasinska
et al., on the other hand, imply $R_V=2.63$ (using the relations provided
by \citealt{Nandy75}) and finally
\begin{equation}
A_{V \,{\rm radio}}=2.10 \, C_{\rm rad} \qquad (\mbox{lower reddening for
distant PN}).
\end{equation}
For all our objects, $C_{\rm opt} > C_{\rm rad}$ (values taken from Stasinska
et al. and \citealt{Tylenda92}), and we have used both the standard reddening
to obtain $A_{V \,{\rm optical}}$ and the lower value $A_{V \,{\rm radio}}$
in our further investigations to estimate a maximum and minimum effect.
Moreover, we have used also a linear mean, \mbox{$A_{V \,{\rm mean}} =
0.5~(A_{V \,{\rm optical}} + A_{V \,{\rm radio}})$}, with larger errors to
account for the present uncertainties. In Table \ref{stapar}, we display
the different distances determined by ``Method~1'' resulting from these
three values of $A_{V}$, whereas, for brevity, we display only the solutions
relying on $A_{V \,{\rm mean}}$ for ``Method~2''.

\begin{figure*}
\begin{minipage}{8.8cm}
\resizebox{\hsize}{!}
   {\includegraphics{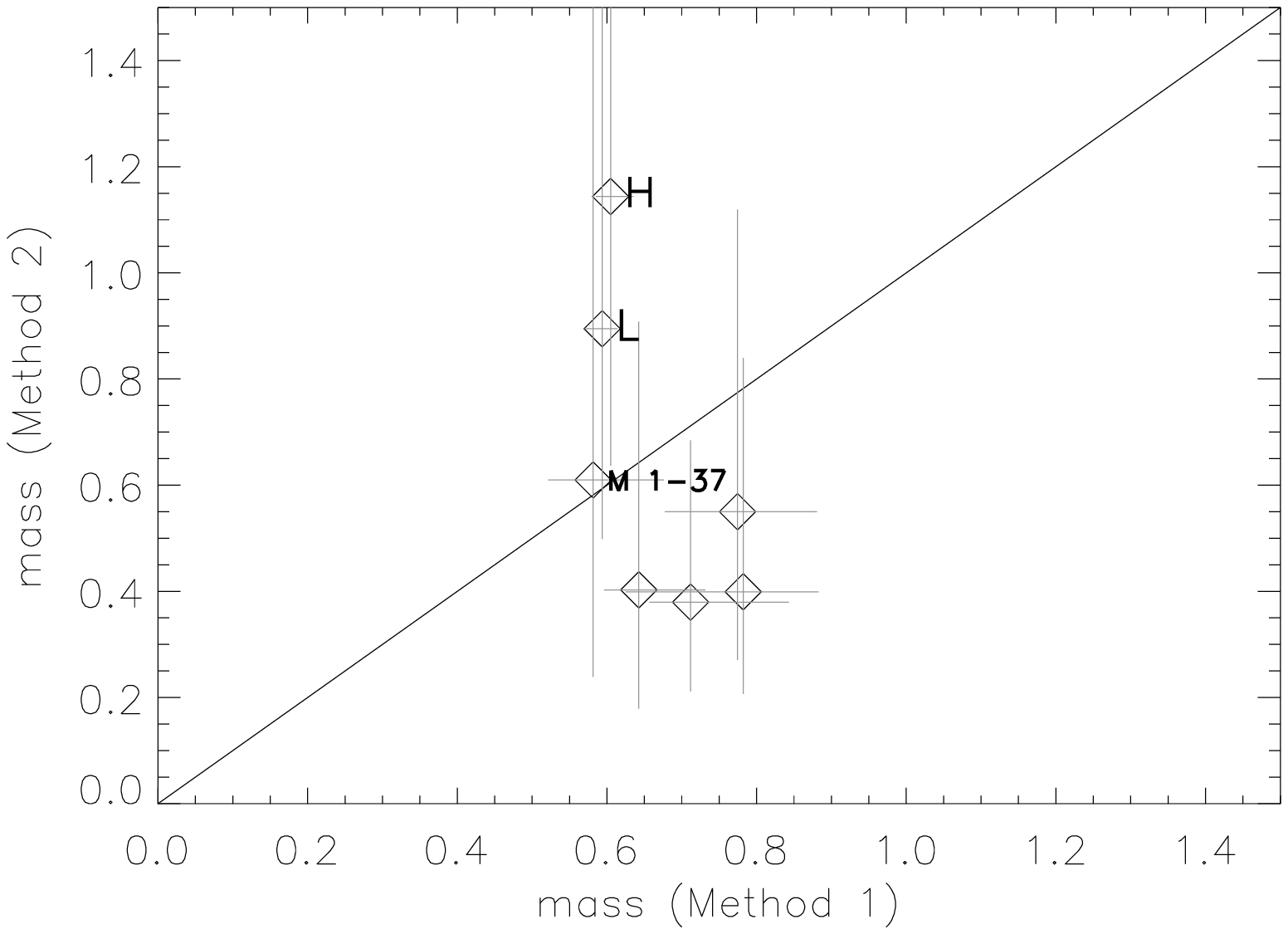}}
\end{minipage}
\hfill
\begin{minipage}{8.8cm}
   \resizebox{\hsize}{!}
   {\includegraphics{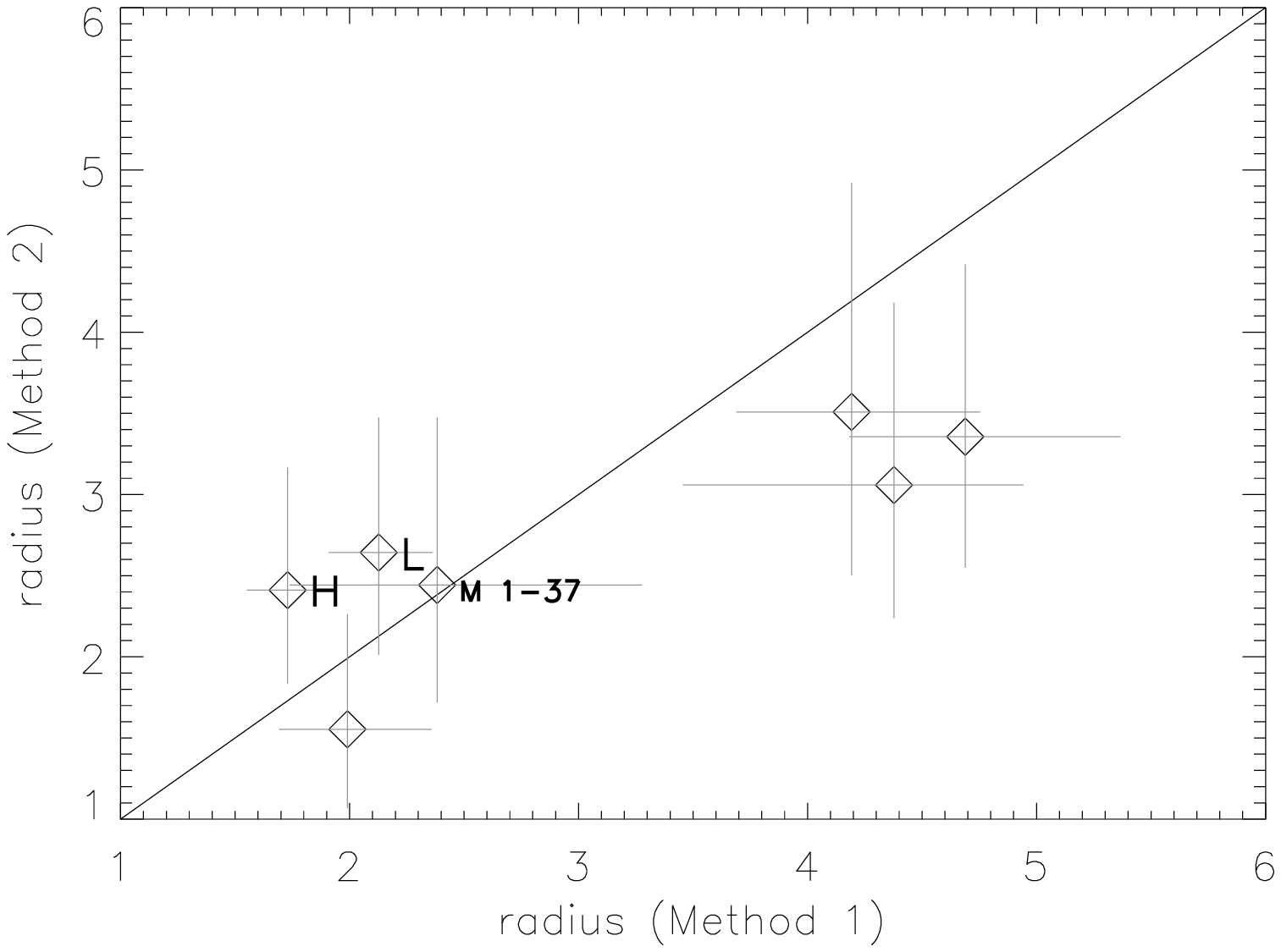}}
\end{minipage}
\\[.3cm]
\begin{minipage}{8.8cm}
\resizebox{\hsize}{!}
   {\includegraphics{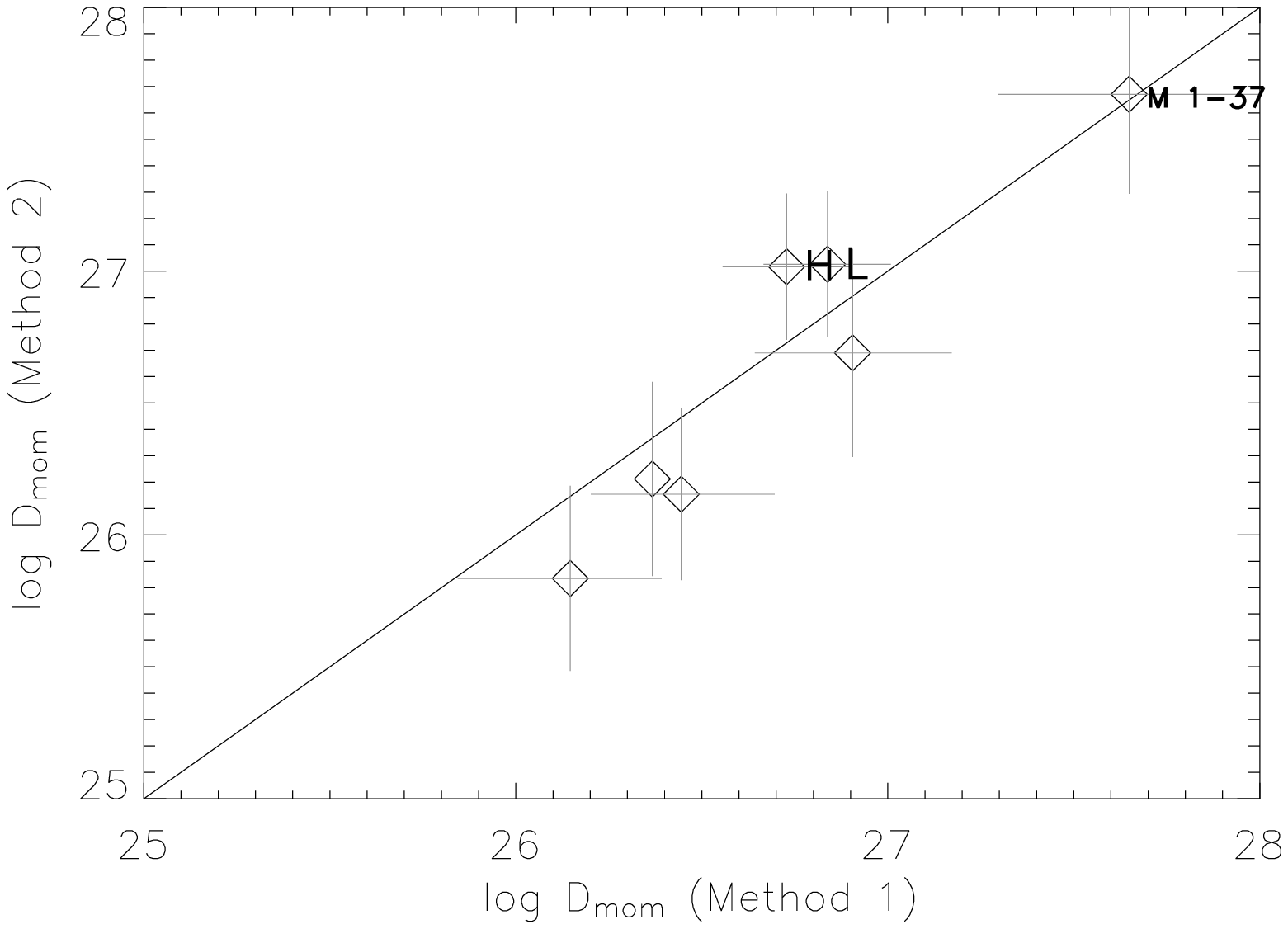}}
\end{minipage}
\hfill
\begin{minipage}{8.8cm}
   \resizebox{\hsize}{!}
   {\includegraphics{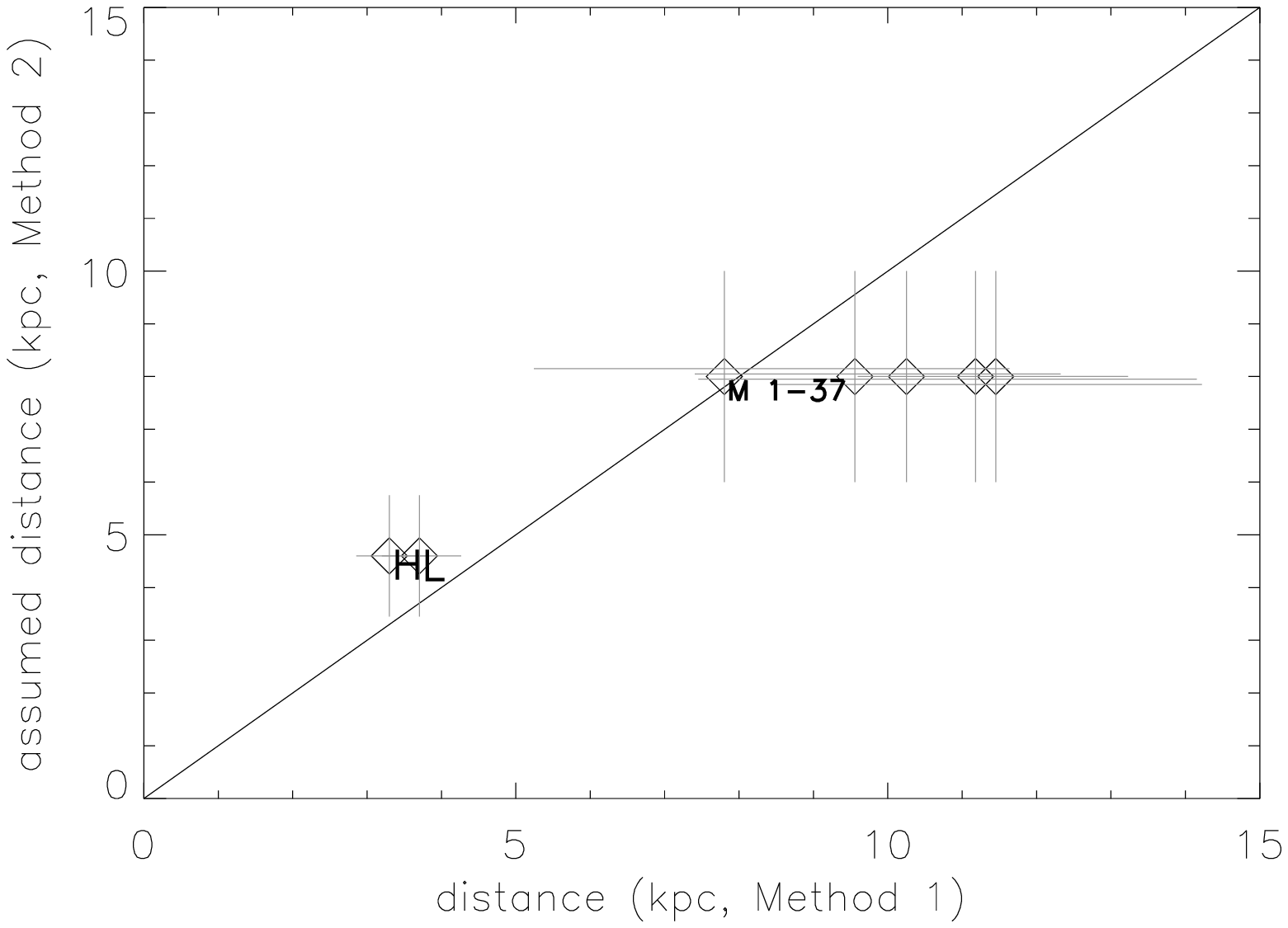}}
\end{minipage}
\caption{Comparison of various parameters of our CSPN sample, derived by
``Method~1'' (based on evolutionary tracks) and ``Method~2'' (based on
adopted distances), using the ``mean'' extinction, $A_{V \,{\rm mean}}$ (see
text): masses in units of $\Msun$, radii in $\Rsun$, modified wind-momenta in
cgs and distances in kpc. $L$ and $H$ refer to the low and high temperature
solution for \Htoe, respectively. The solid line displays the one-to-one
relation. Note that most of the results agree within the error bars, but that
the errors inherent to ``Method~1'' are considerably lower than those relying
on ``Method~2''. Since the gravity of \Mots\ has been adopted from distance
arguments, all parameters agree by definition.} \label{comp} \end{figure*}

\subsection{Comparison of Method~1 and 2: stellar parameters and distances}

Regarding the error propagation of the quantities observed or determined
directly from the spectra (see first part of Table \ref{stapar}) into
deduced physical quantities such as radius, mass, luminosity and mass-loss
dependent quantities we refer the reader to the detailed discussion provided
by \cite{Markova04} and \cite{Repolust04}.

\subsubsection{Mass and radius related quantities}

Figure \ref{comp} (upper left panel) shows a comparison of the derived masses
from the two methods. Obviously, the error bars from ``Method~1'' are much
smaller than those resulting from ``Method~2'', which of course is related
to the fact that ``Method~2'' involves the rather large uncertainties in
$\Rstar$ (due to uncertainties in distance and extinction), but, even more,
because $\Mstar$ depends quadratically on radius. In contrast, for masses
based on evolutionary tracks the dependence on errors in $\lg$ and $\Teff$
should be minor since the slope of the tracks is fairly parallel to the
$\Teff$-$\lg$ iso-contours of equivalent width for the Balmer lines formed
in the photosphere.

Relying on this notion, we defined the mass uncertainty resulting from the
adopted errors in $\lg$ and $\Teff$ in the following way: We determine
$\Mstar$ for the pair ($\lg + \Delta \lg$, $\Teff$) as well as for the pair
($\lg$, $\Teff + \Delta \Teff$), and use the higher of both masses as our
upper limit for $\Mstar$. The lower limit is defined analogously, with
opposite signs. We do {\it not} use combinations of ($+\Delta \lg$, $-\Delta
\Teff$) and vice versa, since such combinations can be excluded from the
results of our analysis (remember that the errors in $\Teff$ and $\lg$ are
not independent).

The resulting uncertainties for $\Mstar$ are amazingly small; even the object
with very large uncertainties in $\lg$, \Mots, has a fairly well determined
mass, which is also related to the fact that for lower masses the dependence
of $\Mstar$ on gravity is rather low. The mass range within our sample
extends from 0.58 to 0.78 $\Msun$, with errors of about 15\%.

From these results, however, one might also argue that our sample is not
completely representative. This is best seen from the fact that the mean
stellar mass of our sample is about $0.68 \Msun$, which is considerably
larger than the expected $0.6 \Msun$ (see, e.g., \citealt{Liebert05}). Such
a discrepancy might be due to selection effects, because for our analysis
we choose only objects with a high enough signal to noise ratio. With such
a bias towards more luminous objects, we might have selected also the more
massive ones.

In contrast, using ``Method~2'', the masses are determined from the radius
which enters quadratically and is dominated by the large uncertainties in
distance and extinction or visual magnitude. If we consider the corresponding
errors as already discussed and summarized in Table~\ref{stapar}, we find
errors in between 30\% to 45\% for $\Rstar$ and uncertainties of the order of
a factor of two (larger or smaller) with respect to $\Mstar$.

\begin{figure*}
\begin{minipage}{8.8cm}
\resizebox{\hsize}{!}
   {\includegraphics{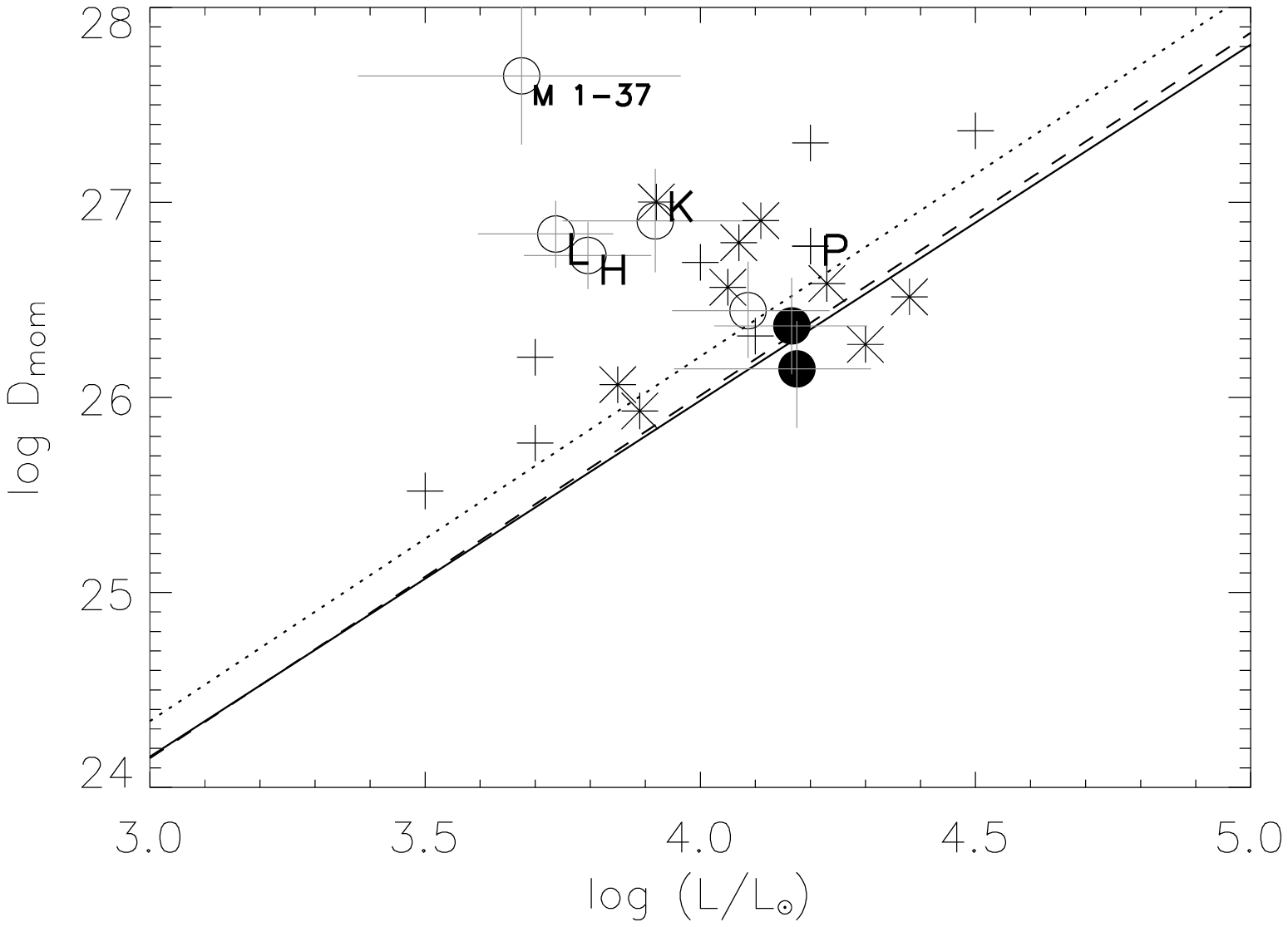}}
\end{minipage}
\hfill
\begin{minipage}{8.8cm}
   \resizebox{\hsize}{!}
   {\includegraphics{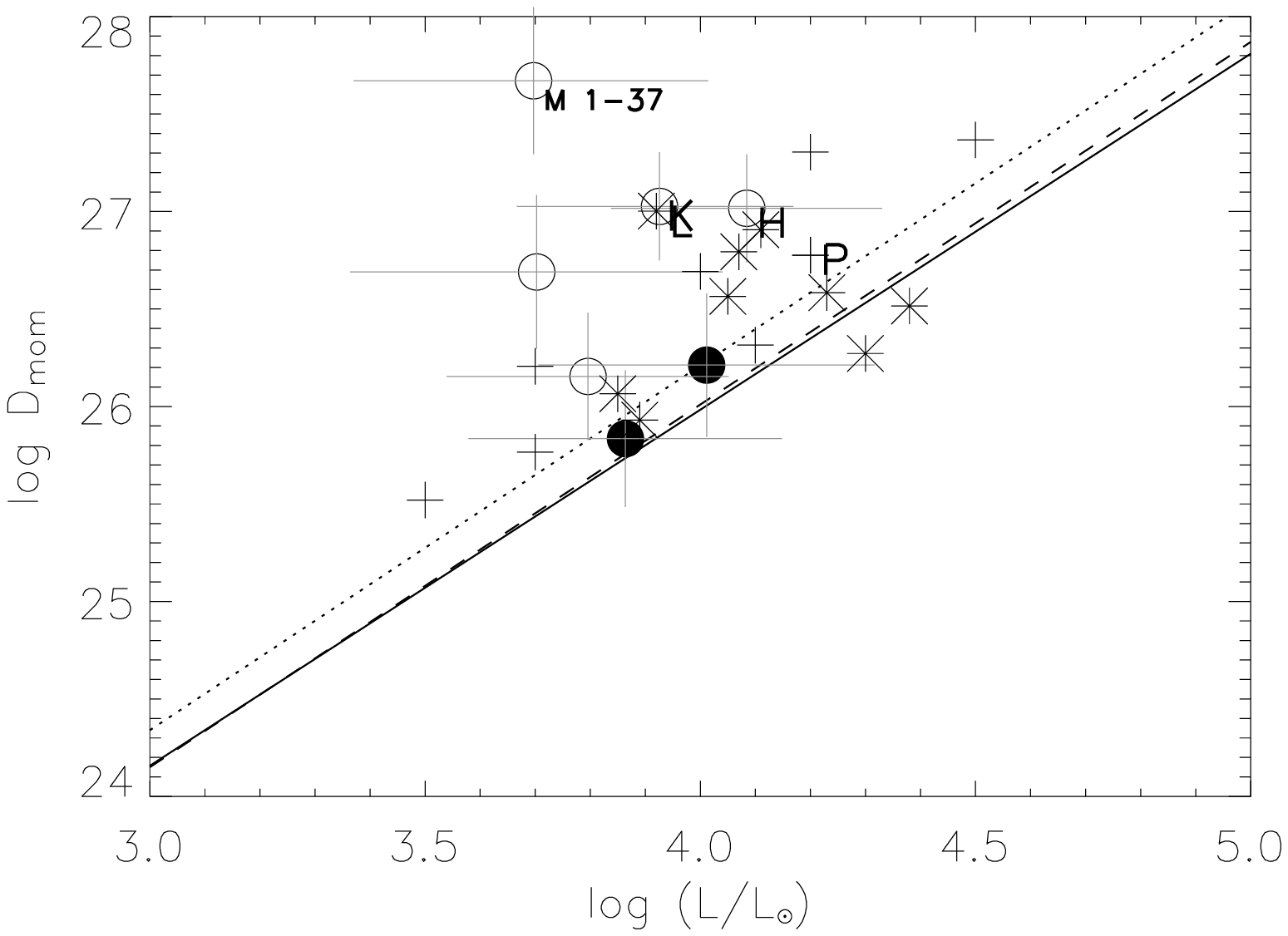}}
\end{minipage}
\caption{Modified wind-momentum rates, $D_{\rm mom}$, as a function of
luminosity. Left: results from ``Method~1''; right: results from
``Method~2''. Circles correspond to the results for our sample (open: $\fcl$
= 1, filled: $\fcl \ne$ 1), asterisks to the results from \citet{Kudritzki06}
and pluses to the results from \citet{Pauldrach04}. The special symbols
denote the various solutions for \Htoe, as in Figure~\ref{stellar_evo}. The
solid line displays the theoretical prediction from \citet{Vink00}, and the
dotted and dashed line the ``observed'' WLR for Galactic O-type supergiants
and giants/dwarfs, respectively (from \citealt{Markova04}). The object
that deviates most from the predicted WLR is the most extreme object in
our sample, \Mots.} \label{wlr} \end{figure*}

Comparing the luminosities shows a similar result. Except for \Mots,
the errors in $\log \Lstar$ derived by ``Method~2'' are twice as large
as those from ``Method~1'', whereas in the case of \Mots\ the errors are
similar, due to the extremely large uncertainties in $\lg$ affecting the
corresponding stellar radius derived by ``Method~1''.

Concentrating on the parameters themselves (mass and radius),
Figure~\ref{comp} shows that for most of the objects the derived values
agree within the error bars, but that the specific values can be quite
different.  Notably, the masses of three bulge-objects as derived from
``Method~2'' would be extremely low, whereas both solutions for \Htoe\
result in implausibly high masses. Again, this is the outcome of the
quadratic influence of a highly insecure stellar radius.

\subsubsection{Distances}

Provided that our atmospheric models do not contain intrinsic errors and
that the evolutionary tracks of the central stars are generally applicable
to all objects of this group, the uncertainties in the spectroscopic
distances are dominated by the errors in the extinction towards the central
stars (plus the error in $V$ for \Mttt). Remember that the derived
masses are very accurate in this case, and that the errors in stellar
radius result from a fairly small error propagation since $\Rstar \propto
\sqrt{\Mstar/g}$. In order to account for the unclear situation regarding
reddening, we assumed rather large errors for $A_{V \,{\rm optical}}$
(standard reddening) and $A_{V \,{\rm radio}}$ ($R_V \approx 2.6$), namely
the mean difference between both quantities for our sample ($\Delta A_V =
\pm 0.6$ mag), whereas for the ``mean'' extinction, $A_{V \,{\rm mean}}$,
we assumed, because of the applied compromise, individual errors of $\Delta
A_{V \,{\rm mean}} = \pm 0.5~(A_{V \,{\rm radio}} - A_{V \,{\rm optical}})$
which are roughly a factor of two smaller.

The resulting distances for $A_{V \,{\rm mean}}$ are displayed in
Figure~\ref{comp} (lower right panel), and those using either $A_{V \,{\rm
optical}}$ or $A_{V \,{\rm radio}}$ are tabulated in Table~\ref{stapar}. At
first note that for the supposed bulge objects we derive distances in the
range of $7.8\,\dots 11.5$\,kpc when using the ``mean'' extinction, whereas
for the alternative solutions $A_{V \,{\rm optical}}$ and $A_{V \,{\rm
radio}}$ we derive distances that are smaller and larger, respectively.

Regarding the specific values, the distances derived by using $A_{V\,{\rm
optical}}$ are best compatible with a bulge population. This becomes
immediately clear if we assume that our sample is statistically distributed,
and we calculate an (error-weighted) mean distance (leaving out \Mots\ since
it has been {\it placed} inside the bulge, see Section~\ref{mots}). In this
case, we obtain the values found in Table~\ref{meandist}. Further
consequences of these results are discussed in Section~\ref{discussion}.

\begin{table}
\begin{center}
\caption{Derived mean distances using different $A_V$ disregarding object
\Mots\ (see text).}
\tabcolsep=6pt
\renewcommand{\arraystretch} {1.3}
\begin{tabular*}{\columnwidth}{l@{\extracolsep\fill}|ccc}
\hline \hline
extinction & $A_{V\,{\rm mean}}$ & $A_{V\,{\rm optical}}$ & $A_{V\,{\rm
radio}}$ \\
\hline
mean distance (kpc) ~ & 10.7 $\pm$ 1.2 & 9.0 $\pm$ 1.6 & 12.2 $\pm$ 2.1 \\
\hline \hline
\end{tabular*}
\label{meandist}
\end{center}
\end{table}

Both distances that we obtain for \Htoe\ (low and high temperature solution)
are fairly similar, and the differences with respect to the re-scaled
distances from \citet{Mendez92} and \citet{Kudritzki06} (compare with the
entry according to $A_{V \,{\rm mean}}$) are due to the different gravity
resulting from our analysis and the applied centrifugal correction, which has
been neglected previously. For this object, the influence of different
extinction/reddening-laws is not as large as for the bulge objects, because
the extinction itself is low.

\subsection{Mass loss rates and wind momenta}
\label{mdotdmom}

Since the diagnostics used (\Ha\ and \HeII\ $\lambda$4686) constrain the
wind-strength parameter, $Q$ (or, even worse, only the product $Q \sqrt
\fcl$), also the mass-loss rates and the (modified) wind momentum rates
depend on radius (and $\vinf$), via $\Mdot = Q ~ (\Rstar ~ \vinf)^{3/2}$
and $\Dmom = Q ~ \Rstar^2 ~ \vinf^{5/2}$. Figure~\ref{comp} (lower left
panel) compares the latter quantities as resulting from ``Method~1'' and
``2''. At least on the logarithmic scale, they are quite similar, though
the errors from ``Method~2'' are larger, of course.

Figure~\ref{wlr} displays our results together with those derived from other
investigations and theoretical predictions, in form of the wind-momentum
luminosity diagram (left panel: ``Method~1''; right panel: ``Method~2'').
Circles correspond to our investigation, denoted as open for objects with
$\fcl$ = 1, and filled for those two objects which show clear evidence
for clumping.

The positions found for our objects scatter in a similar way around
the ``observed'' WLR for Galactic O-type supergiants and giants/dwarfs
(dotted/dashed, from \citealt{Markova04}) and the corresponding theoretical
predictions (solid, from \citealt{Vink00}) as the results obtained by \citet[
asterisks]{Kudritzki06} for a different sample, with \Htoe\ in common. We
also show the results from \citet[ crosses]{Pauldrach04}, obtained from
a UV-analysis based on a consistent solution of the hydrodynamics.

Leaving aside the most extreme outlier\footnote{corresponding to \Mots\
(uppermost left circle), which is also the most extreme object in our sample
(with very strong \SiIII\ lines) and might be in the transition phase
to a WR type object, in which case it {\it must} have a larger $\Mdot$
and wind-momentum rate than ``normal'' CSPN.}, those two CSPN {\it with}
clumping corrections lie close to but below the O-star WLR, one object is
on the WLR and the other two well above. Since we are not able to exclude
clumping corrections for the latter objects, it might be that they also
follow the O-star trend. As already outlined by \citet{Kudritzki06},
however, a definite statement regarding the question of whether CSPN
are a low luminosity extension of the O-star relation (as suggested by
\citealt{pau88ea} and \citealt{Kudritzki97}) or whether there is no such
clear relation (see also \citealt{Tinkler02}) is still not possible at
the present stage.

There is one additional argument which might support the former hypothesis.
If one claims that the CSPN indeed form such an extension, one obtains a
certain constraint on their wind-strength parameter, $\logQ$: Rewritten
in terms of this parameter, the WLR for a certain class of objects reads
\begin{eqnarray}
\log \Dmom &\approx& x \log L + D, \qquad {\rm i.e.,} \nonumber\\
\log Q + 2 \log \Rstar + \frac{5}{2} \log \vinf &\approx& 4\,x \log \Teff +
2\,x \log \Rstar + D'  \nonumber \\
(\vinf &\propto& v_{\rm esc} =  \sqrt{g\Rstar}) \nonumber \\
\log Q + (2  - 2\,x + \frac{5}{4}) \log \Rstar &\approx&  4\,x \log \Teff -
\frac{5}{4} \lg + D'',
\end{eqnarray}
with $x$ and $D$ the slope and offset of the specific WLR. If now the CSPN
would follow such a WLR and were an extension of the O-star relation, this
equation implies
\begin{equation}
(\log Q - 0.5 \log \Rstar)_{\rm CSPN} = (\log Q - 0.5 \log \Rstar)_{\rm
O-stars}\,,
\end{equation}
if we compare CSPN and O-stars of similar effective temperatures and
gravities, using a typical slope of $1/x=0.53$ for the O-star WLR. In other
words, the average wind-strengths of CSPN must be roughly 0.5 dex larger than
the corresponding O-star values at same $\Teff$ and $\lg$, if we adopt a
typical difference of 1 dex in radii. But this is what we actually {\it see}
in our spectra (cf. Section~\ref{comments}): Compared to O-supergiants of
similar spectral type, all our sample stars show much more emission in \Ha,
immediately pointing to a higher $\log Q$. Indeed, the average value for our
sample is $\logQ$ = $-$11.9, compared with typical (late) O-supergiant values
in between $-$12.8 to $-$12.4.

Summarizing, a similar WLR implies that for CSPN one expects larger $Q$
(for smaller $\Rstar$ and $\vinf$), whereas, vice versa, the observed,
strong \Ha\ emission might imply that the CSPN indeed lie on the O-star WLR.
Alternatively, there is the possibility that the clumping properties of
O-stars and CSPN differ significantly, with stronger clumping in CSPN.

\section{Discussion}
\label{discussion}

This project has produced a remarkable mixture of encouraging and alarming
results. On the positive side, leaving aside \Mots\ with its unique spectrum
and uncertain $\lg$, the four remaining bulge central stars turn out to have
``Method~1'' spectroscopic distances within $\pm 10$\% to $\pm 15$\%
(depending on the extinction law) of their average distance, well within the
expected uncertainties. The alarming fact is of course that their average
distance, if we adopt the reddening corrections suggested by
\citet{Stasinska92}, is a factor 1.5 larger than the currently accepted
distance to the Galactic bulge. This allows the following possible
interpretations:

\begin{itemize}
\item[(1)] If the reddening corrections based on a low $R_V$ and radio-\Hb\
extinctions are correct, then the combination of our model atmospheres plus
standard post-AGB evolutionary tracks gives systematically too large
spectroscopic distances, despite their excellent internal agreement.
\item[(2)] If the standard reddening law is valid, and using $C_{\rm opt}$,
then our ``Method~1'' spectroscopic distances give the right distance to the
Galactic bulge to within 13\%, which would have to be considered a successful
test; but then we need to find a reason for the observed systematic
difference between the optical and radio-\Hb\ extinctions.
\end{itemize}
In the present situation we do not feel that we have enough information to
decide between (1) and (2). Our referee, A. Zijlstra, has pointed out
that our sample objects have (by bad luck) relatively poorly known radio
fluxes. Most come from very old radio data, obtained at the very start of
the VLA, and we cannot be as confident of these as we are of more recent
observations. Only \Htts\ can be considered as reliable; the others need
confirmation.

The analysis of our comparison object from the ``solar neighborhood'',
\Htoe, reveals a prototypical CSPN mass, of about $0.6 \Msun$ (both for the
low and high $\Teff$ solution) if ``Method 1'' is followed. This is in
agreement with the $0.67 \Msun$ derived by \citet{Mendez92} and the recent
value of $0.63 \Msun$ by \citet{Kudritzki06}. The higher masses derived using
``Method 2'', however, lacks any significance (except for the fact that it
overlaps within the error bars with our ``Method 1''-mass), due to its highly
uncertain distance (adopted from similar analyses) and the large error
propagation of this quantity. Recently, \citet{Napiwotzki06} has claimed,
using an analysis of kinematical parameters based on simulated Galactic
orbits, i.e., by means of a completely different approach, that a low mass
similar to our ``Method 1'' result would be more compatible with the
kinematical parameters one would expect from a member of the thin-disk
population.

On the other hand, \citet{Pauldrach04} derived a significantly larger mass
of $1.33 \Msun$ for this object. In contrast to our approach, they have
solved the hydrodynamical equations in a self-consistent way, and compared
the corresponding, synthetic UV spectra with observations. $\Teff$ then
follows from the observed ionization equilibrium, whereas mass and radius
are uniquely constrained from the terminal velocity (being proportional to
the photospheric escape velocity, $\propto (\Mstar/\Rstar)^{1/2}$, and
mass-loss rate (depending on luminosity, i.e., again on stellar radius).
Note that both wind parameters are clearly ``visible'' in the UV-lines.

Ideally, the results of optical and UV spectroscopy should be similar. By
comparing the derived effective temperatures and gravities for our high
$\Teff$ solution (Fig.~\ref{stellar_evo}, `H' vs. `P'), at least for these
parameters this is indeed the case (i.e., the ratios $\Mstar/\Rstar^2$ do
agree). Thus, the difference in mass is due to a different radius
(implying different ratios $\Mstar/\Rstar$) which in the case of
\citet{Pauldrach04} results from hydrodynamical considerations, whereas here
we have derived the mass directly from evolutionary tracks.

A solution of this discrepancy cannot yet been provided, but will be
commented on in the following. At first, let us give a few additional remarks
concerning PN distances. In the same publication as just referred to,
\citet{Pauldrach04} presented a comparison between spectroscopic and other PN
distance determinations, which will not be repeated here except to comment on
the fact that one of the very few cases of a clear discrepancy (Hipparcos
parallax of PHL~932) has been resolved in favor of the spectroscopic
distance. \citet{Harris07} reports new trigonometric parallaxes for several
central stars, pointing out that the Hipparcos parallax of PHL~932 is wrong.
The new geometric distance of PHL 932 is 300\,pc, in perfect agreement with
the spectroscopic distance we obtain if we assume a stellar mass of
0.3\,$\Msun$ and the following atmospheric parameters: $\Teff = 36\,000$\,K
and $\lg = 5.7$, which are the averages from two independent spectroscopic
studies (\citealt{Mendez88b} and \citealt{Napiwotzki99}). Clearly the Harris
geometric distance confirms the small stellar mass and proves definitely that
PHL~932 cannot be a post-AGB star \citep{Mendez88b}. So here we have a case
where the model atmospheres pass the test. \citet{Harris07} also shows
agreement, within the uncertainties, between his parallaxes and spectroscopic
distances for several PN, mostly from \citet{Napiwotzki99}.

Of course we must add that those central stars with trigonometric parallaxes
are high-gravity stars with no wind features; so this agreement is not a
confirmation of the wind models. The same comment applies to the central star
of K~648, the PN in the globular cluster M~15, whose spectroscopic distance
is in excellent agreement with the cluster distance \citep{McCarthy97}. Let
us argue in the following way: if the combination of post-AGB theory and
windless model atmospheres gives good distances (as indicated by the
high-gravity stars and K~648) and the combination of post-AGB theory and wind
models fails systematically, then it would be easy to conclude that the wind
models must be wrong. But this is not the case -- evidence for such
systematical failures have not been found so far, and it should be noted that
our Galactic bulge sample includes central stars with {\it weak and strong}
winds, yet all their ``Method~1'' spectroscopic distances agree, with no
differences that could be attributed to wind strength effects. Alternatively,
we have the possibility that another group of CSPN with winds exists (e.g.,
some objects from the sample analyzed by \citealt{Pauldrach04}), which follow
a different evolutionary path and for which the standard CSPN tracks do not
apply. A further indication of such a possible qualitative difference is the
fact that \citet{Kudritzki06}, using a similar method as in this paper, also
based on the standard evolutionary tracks, found higher masses (around
$0.74\,\Msun$) for the same sample of objects as analyzed by
\citet{Pauldrach04}, whereas the mean mass of the CSPN in our current sample
is around $0.68\,\Msun$.

In summary, the combination of excellent internal agreement (and some
additional supporting evidence) plus the extinction problem leads us to
conclude that our Galactic bulge distance test is undecided.\\

\noindent We can see two ways of trying to force a decision:
\begin{itemize}
\item[(1)] a careful, multi-wavelength redetermination of the interstellar
extinction toward the bulge PN we have studied,
\item[(2)] enlarge the sample of bulge central stars, trying to get
high-quality, high-resolution spectrograms with minimum nebular
contamination.
\end{itemize}
Alternatively, as already emphasized in the introduction, post-AGB tracks and
model atmospheres could be {\it independently}\/ tested as soon as
medium-resolution spectroscopy of CSPN in the Magellanic Clouds, with
efficient nebular light subtraction, becomes feasible; the advantage being
the small amount of reddening in that direction.

\section{Summary and conclusions}
\label{summary}

We have selected a small sample of PN central stars in the Galactic bulge
and obtained high-resolution spectrograms with the Keck HIRES spectrograph.
We present spectral classification of these central stars and we model
their optical spectral features (including wind effects) using a state
of the art non-LTE, spherically symmetric model atmosphere code that
includes the effects of line blocking/blanketing and wind clumping. We
obtain effective temperatures, surface gravities, wind parameters and He
and Si abundances.  With one exception, He and Si abundances are normal
for H-rich central stars.  The exception is \Mots, a central star with
spectacular P Cygni profiles indicative of a very dense wind, He-rich,
and with an apparently very high Si abundance, 28 times solar, unexplained
for the moment. This star also shows strong unidentified emissions at 4485
and 4504 \AA, which we have refrained from identifying with Silicon.

Armed with the stellar atmospheric parameters we then calculate other
parameters not directly derivable from our analysis, using two methods.
``Method~1'' uses post-AGB evolutionary tracks to read masses, from which
stellar radii, luminosities, mass loss rates and spectroscopic distances
follow. ``Method~2'' assumes the distance to the Galactic bulge and derives
stellar radii and other quantities; this method turns out to be very
uncertain because of the strong dependence of mass on stellar radius, which
is itself not well determined.

If we focus our attention on the results of ``Method~1'', the most remarkable
conclusion regards the internal agreement of the spectroscopic distances of
at least four bulge central stars within our sample (leaving aside the
extreme object \Mots\ with its ill-defined gravity): all of them are quite
the same to within $\pm 10$\% to $\pm 15$\% (depending on the extinction law)
of their average distance. This result is well within the expected
uncertainties. There is a problem, however: the average distance to the
Galactic bulge is strongly dependent on the adopted reddening correction, to
such an extent that we cannot decide if ``Method~1'' passes the consistency
test for our sample (if we adopt the most recently derived reddening
corrections, the average distance turns out to be a factor 1.5 larger than
the currently accepted distance to the Galactic bulge). Additional studies of
extinction for the PN in our sample will be necessary, and it would be good
to increase the size of the bulge CSPN sample subject to this kind of
analysis, while we wait for a chance to repeat this kind of study in the
Magellanic Clouds, which are almost free of the extinction problem.

\begin{acknowledgements}
We like to thank our referee, A. Zijlstra, for his useful comments
and suggestions on the first version of the manuscript.
This work has been supported by the Deutsche For\-schungs\-ge\-mein\-schaft
under grant PA~477/4-1 and the ``Son\-der\-for\-schungs\-be\-reich
\mbox{375} f\"{u}r Astro-Teil\-chen\-phy\-sik''. P. Hultzsch would also
like to thank the Max-Planck Ge\-sell\-schaft for additional funding.
\end{acknowledgements}

\bibliographystyle{aa}
\bibliography{cspn}
\end{document}